\documentclass[12pt]{article}
\usepackage{amsmath}
\usepackage{graphicx,psfrag,epsf}
\usepackage{enumerate}
\usepackage{natbib}

\usepackage{url}
\usepackage{algorithm}
\usepackage{algorithmic}

\newcommand{\blind}{1}

\begin{document}

\def\spacingset#1{\renewcommand{\baselinestretch}%
{#1}\small\normalsize} \spacingset{1}


\if1\blind
{
  \title{\bf A Bayesian Framework for Non-Collapsible Models}
  \author{Sepehr Akhavan Masouleh, Babak Shahbaba, \\and Daniel L. Gillen \\
    Department of Statistics, University of California, Irvine}
  \maketitle
} \fi

\if0\blind
{
  \bigskip
  \bigskip
  \bigskip
  \begin{center}
    {\LARGE\bf A Bayesian Framework for Non-Collapsible Models}
\end{center}
  \medskip
} \fi

\bigskip

\begin{abstract}
{In this paper, we discuss the non-collapsibility concept and propose a new approach based on Dirichlet process mixtures to estimate the conditional effect of covariates in non-collapsible models. Using synthetic data, we evaluate the performance of our proposed method and examine its sensitivity under different settings. We also apply our method to real data on access failure among hemodialysis patients.}
\end{abstract}

\noindent%
{\it Keywords:} Dirichlet process mixture models; Survival analysis; Generalized linear models
\vfill

\newpage
\spacingset{1.45} 

\section{Introduction}
\label{nonCollapIntro}

Statistically, non-collapsibility represents the setting where the marginal measure of association between two random variables $X$ and $Y$, differs from the conditional measure of association between these two random variables, after conditioning upon the levels of a third random variable $Z$, where $Z$ is not a confounder, i.e., $Z$ is associated with one random variable but not the other \citep{greenland1999confounding}. In this situation, a careful attention is required to properly interpret a conditional association as opposed to a marginal association. Further, one should note that in the absence of confounding, both the marginal association and the conditional association, despite being different, are unbiased. Hence, a clear distinction between confounding and non-collapsibility is required.

Similarly, non-collapsibility exists in a regression setting when the marginal association between a predictor variable, $X$, and a response variable, $Y$, differs from the conditional association in a separate regression model where a third variable $Z$ is adjusted in the model. As before, we assume that $Z$ is not a confounder so it is only associated with the response variable. 

In general, one needs to consider the relative importance of estimating the marginal association between the two random variables $X$ and $Y$, in contrast to the conditional association given a third random variable $Z$.  When $Z$ is observed, it is possible to heuristically compare the difference between the marginal and the conditional associations by simply comparing the adjusted and unadjusted estimated associations.  However, when $Z$ is latent, analysts generally default to estimating a marginal association without giving any thoughts to the relative merits of the two estimands. 

In longitudinal studies, non-collapsibility has especially garnered some attention when comparing the estimates from the generalized linear mixed model with the estimates from the generalized estimating equation model, where the former provides conditional estimates that are conditioned upon the subject-specific random effects, and the latter provides estimates that are marginalized over all subjects. Longitudinal data can be considered as a special case of the repeated measure data with measurements indexed by time. We shall use the words ``longitudinal data" and ``repeated measure data" interchangeably.

As a simple case, one may consider $n$ subjects, each with $l_i$ within subject measurements with $Y_{ij}$ and $t_{ij}$ as the outcome and the covariate for the $j^{th}$ measurement on the $i^{th}$ subject, respectively. One can write a generalized linear mixed effect model with random intercepts of the form 
\begin{align*}
E[Y_{ij}| t_{ij}, \beta_{0i}] &= \mu_{ij},
\end{align*}
where the mean $\mu_{ij}$ and the covariate $t_{ij}$ and the subject-specific random intercept $\beta_{0i}$ are linked using a link function $g(.)$, where
\begin{align}
g(\mu_{ij}) &= \beta_{0i} + \beta_0 + \beta_1 t_{ij}. \label{Ch3CondGLMMmodel}
\end{align}
In this model, $\beta_0$ and $\beta_1$ are intercept and slope that are shared across all subjects. In a typical mixed effects model, $\beta_{0i}$, where $i \in\{1, \dots, n\}$, are assumed to be independent and Normally distributed. Under this model setting, conditioned upon the subject-specific random intercepts, $\beta_{0i}$, $\beta_1$ represents the conditional association between the random variable $t$ and the outcome, $Y$.

Alternatively, one may consider a model of the form 
\begin{align*}
E[Y_{ij}| t_{ij}] &= \eta_{ij},
\end{align*}
where the mean $\eta_{ij}$ is related to the covariate $t_ij$ through a link function $g(.)$, where
\begin{align}
g(\eta_{ij}) &= \gamma_0 + \gamma_1 t_{ij}.\label{Ch3MargGLMMmodel}
\end{align}
In this model, $\gamma_0$ is the intercept and $\gamma_1$ is the slope where both are shared across all subjects. Under this model setting, $\gamma_1$ represents the marginal association between the covariate $t$ and the outcome, $Y$.

Generally, even with random intercepts with no confounding effect, the conditional covariate effect $\beta_1$ (equation (\ref{Ch3CondGLMMmodel})) and the marginal covariate effect $\gamma_1$ (equation (\ref{Ch3MargGLMMmodel})) need not be equal. Several authors including \cite{gail1984biased}, \cite{gail1986adjusting} showed that with non-confounding subject-specific random intercept, $\beta_{0i}$, $\beta_1$ is guaranteed to be collapsible, if $g(.)$ is either the identity link or the log link. That means with the identity or the log link and in the absence of confounding, equality of the conditional covariate effect $\beta_1$ and the marginal covariate effect $\gamma_1$ is guaranteed. Hence, we are primarily interested in studying non-collapsibility in logistic and proportional hazards models. 

\begin{figure}[!htb]
\centering
\includegraphics[scale = 1, width=0.75\linewidth]{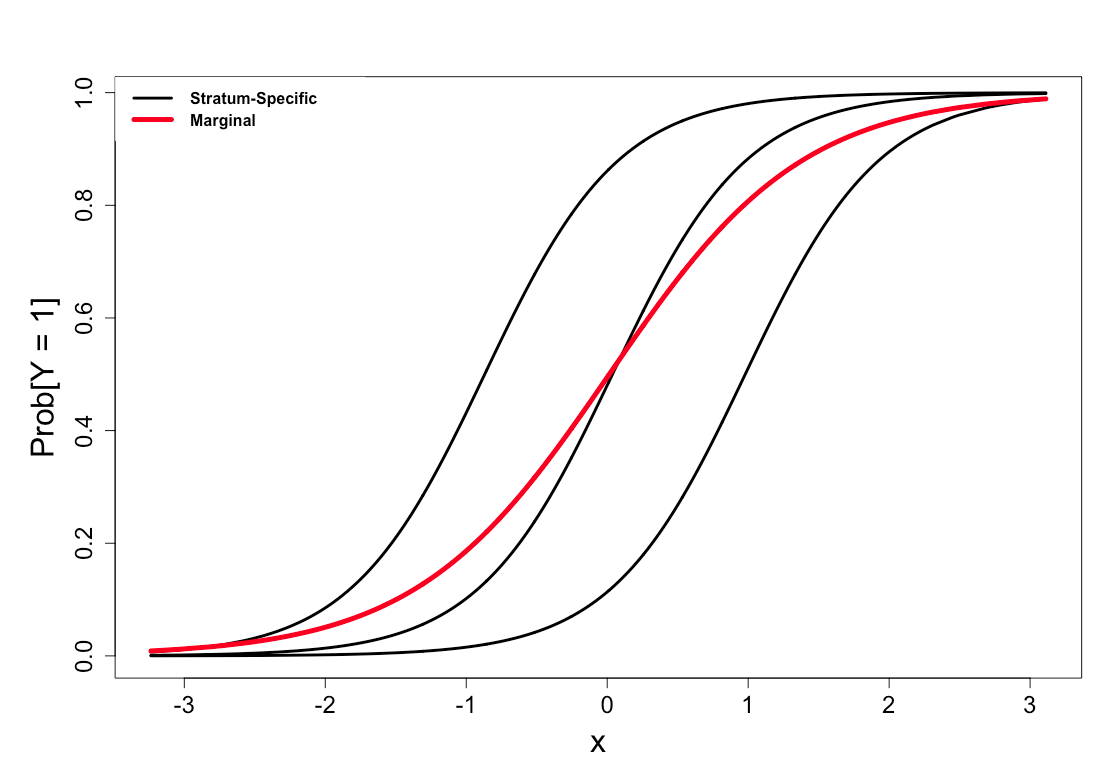}
\caption{Graphical representation of non-collapsibility in logistic regression using synthetic data. Synthetic binary data were generated with three sub-groups with different intercept of $\beta_{01} = -2$, $\beta_{02} = 0$, $\beta_{03} = 2$. Independently of the intercepts, covariate $X$ was simulated from the standard Normal $N(\mu = 0, \sigma^2 = 1)$. This figure shows that the marginal slope (in red) is smaller than the stratum-specific slope (in black).}
\label{LogistNonCol}
\end{figure}

To show the non-collapsibility effect in the logistic regression model, we generated synthetic data, where we considered three different groups with different intercepts of $\beta_{01} = -2$, $\beta_{02} = 0$, $\beta_{03} = 2$. Independently of the intercepts, we generated covariate $X$, where $X$ is simulated from the standard Normal $N(\mu = 0, \sigma^2 = 1)$. Using the a logistic link and with a true coefficient values of  $\beta_1 = 2$, we generated binary outcomes. We then fit a conditional model of the form
\begin{align*}
logit[E(Y_{ij} | X_{ij}, \beta_{0i})] = \beta_{0i} + \beta_0 + \beta_1 X_{ij},
\end{align*}
where $Y_{ij}$ is a binary outcome for the $j^{th}$ measurement on the $i^{th}$ cluster, $X_{ij}$ is the covariate value corresponding to the outcome $Y_{ij}$, and  $\beta_{0i}$ is the true value of the cluster-specific intercept that is directly adjusted in the model. We also fit a marginal model of the form
\begin{align*}
logit[E(Y_{ij} | X_{ij})] = \gamma_0 + \gamma_1  X_{ij}.
\end{align*}
After fitting the conditional and the marginal models above, we plot the results, where the x-axis is the covariate values and the y-axis is the predicted probability of $Y = 1$. In this plot, the red curve shows the predicted values from the marginal model and the three black curves show the predicted values from the conditional each corresponding to a sub-group. As Figure \ref{LogistNonCol} shows, the marginal slope that is averaged across sub-groups ($\gamma_1$) is smaller than the stratum-specific slope ($\beta_1$). This plots clearly shows non-collapsibility in logit link.

As Figure \ref{LogistNonCol} shows, the marginal coefficient estimand $\gamma_1$ is shrunk towards the null hypothesis of no covariate effect compared to the conditional coefficient estimand $\beta_1$. When random intercepts are latent, even under the conditional generalized linear mixed effects model (equation (\ref{Ch3CondGLMMmodel})), the coefficient estimate $\hat{\beta_1}$ may shrink towards 0 compared to the true conditional estimand and that is when the distribution of the random intercepts are mis-specified. One such example is a random intercept model with true random intercepts distributed according to a bi-modal distribution. In this situation, coefficient estimates under a model that assumes random intercepts are distributed Gaussian, may still attenuate towards $0$ compared to the true conditional coefficients. 

Similar to the logistic regression models, proportional hazards models are also non-collapsible. Let $T_{ij}$ denote the $j^{th}$ survival time for the $i^{th}$ cluster. One example of such repeated measure survival data is the survival data on access failure among hemodialysis patients where each patient may have multiple access failures. Let $X_{ij}$ be the covariate corresponding to the $T_{ij}$ survival outcome. One can write a multiplicative hazard function of the form
\begin{align*}
h(T_{ij} | X_{ij}, \beta_{0i}) = h_0(T_{ij}) exp\{\beta_{0i} + \beta_1 X_{ij}\},
\end{align*}
where $h(T_{ij} | X_{ij}, \beta_{0i})$ is the hazard at time $T_{ij}$, $h_0(T_{ij})$ is the baseline hazard at time $T_{ij}$, $exp\{\beta_{0i}\}$ is the frailty term including latent cluster-specific baseline hazard multipliers, and $\beta_1$ is the log relative risk of the effect of the covariate $X_{ij}$ on the risk of ``death". Under this model setting, $\beta_1$ is the conditional relative risk of the covariate $X_{ij}$ that is conditioned on the frailty them $exp\{\beta_{0i}\}$. 

Alternatively, a marginal proportional hazards model is of the form
\begin{align*}
h(T_{ij} | X_{ij}) = h_0(T_{ij}) exp\{\gamma_1 X_{ij}\},
\end{align*}
where $h(T_{ij} | X_{ij})$ and $h_0(T_{ij})$ are hazard and baseline hazard at time $T_{ij}$, respectively. $\gamma_1$ represents the marginal log relative risk of the effect of every one unit changes in the covariate $X_{ij}$ on the risk of ``death". As we shall show with synthetic data, proportional hazards models are non-collapsible where the marginal parameter $\gamma_1$ shrinks toward $0$ compared to the conditional parameter $\beta_1$. 

In this paper, we explore non-collapsibility in longitudinal data when there exists latent subject-specific random intercepts. For non-collapsible logistic regression and proportional hazards models, we propose Dirichlet process mixture models \citep{antoniak74,sethuraman94} that are capable of detecting underlying structure of data by clustering units of analysis into sub-clusters based on the distributional similarities of those units. We believe that our approach can provide insights into conditional associations between the response variable and a set of covariates given population subgroups. Using simulation studies, we compare our proposed models with the common statistical models to analyze longitudinal data. Finally, we use our proposed models to analyze data on hemodialysis patients in order to find risk factors associated with access failure among these patients.

\section{Methods}
\label{nonCollapMethod}

With the focus on logistic regression and the proportional hazards models, and in the context of modeling correlated longitudinal data where repeated measures on sampling units are collected over time, we propose Dirichlet process mixture models capable of estimating conditional covariate effects when there exists latent sub-population effects. In Section \ref{DPMLogisticReg}, we introduce our proposed Bayesian logistic model, and in Section \ref{DPMPH} we introduce our proposed Bayesian proportional hazards model. 

\subsection{A Bayesian Hierarchical Logistic Regression with Dirichlet Process Mixture Priors}\label{DPMLogisticReg}

The logistic link is non-collapsible. This means, when there exists latent population subgroup effects in the form of random intercepts, failure to adjust for these subgroup effects leads to coefficient estimates that are shrunk toward $0$ compared to the true conditional estimands from a separate model with those latent random intercepts taken into account. Generalized linear mixed effects models are capable of modeling random intercepts where they typically assume random intercepts to be distributed according to a Gaussian distribution, however, distributional mis-specification of the random intercepts may still cause coefficient estimates to shrink. A model capable of detecting subgroup random intercepts, that is also robust to distributional mis-specification of random intercepts, can provide the merits of estimating the conditional coefficient estimates.   

We propose a hierarchical Bayesian model that is capable of detecting latent subgroup effects that are in the form of latent random intercepts. The models is capable of estimating conditional parameters. Using a Dirichlet process mixture prior, our proposed model is robust to distributional mis-specification of the random intercepts. In our proposed model, we consider the binary data $Y_{ij}$ to be distributed according to 
\begin{align*}
Y_{ij}|\beta_{0i}, \beta_0, \beta_1, X_{ij} &\sim Bernoulli(p_i = \beta_{0i} + \beta_0 + \beta_1 X_{ij}),
\end{align*}
where $i \in \{1, \dots, n\}$ and $j \in \{1, \dots, l_i\}$ with $n$ as the number of subjects and $l_i$ as the number of measurements on the $i^{th}$ subject, $X_{ij}$ is the corresponding covariate to the outcome $Y_{ij}$, $\beta_{0i}$ is the subject-specific intercept for subject $i$, and $\beta_0$ and $\beta_1$ are the intercept and the slope that are shared across all subjects, respectively. We consider Gaussian priors on the shared intercept $\beta_0$ and the shared slope $\beta_1$ of the form
\begin{align*}
\beta_{0} &\sim N(0, \sigma^{2}_{\beta_{0}}),\\
\beta_{1} &\sim N(0, \sigma^{2}_{\beta_{1}}).
\end{align*}
We propose using the Dirichlet process mixture prior on the random intercepts $\beta_{0i}$, where $i \in \{1, \dots, n\}$ with $n$ as the number of subjects in the data. Using the Dirichlet process mixture prior, as opposed to an explicit distributional assumption, will make the model robust to distributional mis-specification. Further, DPM prior will allow subjects to cluster based on the distributional similarities of their latent random intercepts, hence, provides higher precision in estimating those latent subject effects. We specify a Dirichlet process mixture prior on $\beta_{0i}$ as
\begin{align}
\beta_{0i} & \sim N(\mu_i, \sigma^{2}_{\beta_{0i}}), \nonumber \\
\mu_i|G & \sim G, \label{LogitBayesMeanDPM}\\
G &\sim DP(\alpha, G_0 = N(0, \sigma^{2}_{0})). \nonumber
\end{align}
The Dirichlet process mixture prior above induces a prior on $\beta_{0i}$ that is essentially an infinite mixture of Normal distributions that are mixed over the mean parameter. We shall refer to this model as a Mean-DPM model.
Alternatively, one may set a Dirichlet process prior that induces an infinite Gaussian mixture prior that are mixed over the standard deviation parameter. Such a prior can be specified as
\begin{align}
\beta_{0i} &\sim N(0, \sigma^{2(i)}_{\beta_0^{(i)}}), \nonumber\\ 
\sigma^{2(i)}_{\beta_0^{(i)}}|G & \sim G, \label{LogitBayesSigmaDPM}\\
G & \sim DP(\alpha, G_0 = log-Normal(\mu_{G_0}, \sigma^{2}_{G_0})). \nonumber
\end{align}
We shall refer to this model as Sigma-DPM model. 

\subsection{A Bayesian Hierarchical Proportional Hazards Model with Dirichlet Process Mixture Priors}\label{DPMPH}
Similar to the logistic regression models, proportional hazards models are also non-collapsible. When there exists differential subject-specific baseline hazard risk, even in the absence of confounding in the baseline hazard risks, failure to adjust for these subject-specific baseline risks in a proportional hazards model leads to coefficient estimates that are shrunk toward $0$ compared to the true conditional estimands from a separate model with those latent baseline risks taken into account. In this situation, a proportional hazards model that is capable of detecting subject-specific baseline hazards, can provide the merits of estimating the conditional coefficient estimates.   

We propose a hierarchical Bayesian proportional hazards model that is capable of detecting the differential subject-specific baseline hazard risk across subjects. Our proposed model uses a Dirichlet process mixture prior on the latent subject-specific baseline hazards. The Dirichlet process mixture prior allows clustering subjects based on the distributional similarities of their baseline hazards. Further, by using the Dirichlet process mixture prior, we avoid any explicit distributional assumption on the latent subject-specific baseline hazards. In our proposed model, we consider survival times $T_{ij}$, where $i \in \{1, \dots, n\}$ and $j \in \{1, \dots, l_i\}$ with $n$ as the number of subjects and $l_i$ as the number of measurements on the $i^{th}$ subject, to be distributed according to a Weibull distribution of the form
\begin{align*}
T_{ij} | \tau, \beta_{0i}, \beta_0, \beta_1, X_{ij} & \sim Weibull(\tau, \theta_i), \\
log(\theta_i) &=  \beta_{0i} + \beta_0 + \beta_1 X_{ij},
\end{align*}
where $X_{ij}$ is the covariate value corresponding to the $T_{ij}$ survival time, $exp\{\beta_{0i}\}$ is a subject-specific baseline hazard, $\beta_0$ is a shared intercept across all subjects, $\beta_1$ is a shared slope across all subjects that represents the log relative risk of every one unit increase in the covariate $X_{ij}$, $\tau$ is the shape parameter, and $\theta_i$ is a subject-specific scale parameter. In the model specification above, we introduced covariates into the model through the scale parameter and using the equation $log(\theta_i) =  \beta_{0i} + \beta_0 + \beta_1 X_{ij}$.
For our proposed model, we consider Gaussian priors on $\beta_0$ and $\beta_1$ parameters as
\begin{align*}
\beta_{0} & \sim N(0, \sigma^{2}_{\beta_{0}}),\\
\beta_{1} & \sim N(0, \sigma^{2}_{\beta_{1}}),
\end{align*}
where $\sigma_{\beta_{0}}$ and $\sigma_{\beta_{1}}$ are fixed numbers.
We also assume a log-Normal prior on the shape parameter, $\tau$, as
\begin{align*}
\tau & \sim log-Normal(\mu_{\tau}, \sigma^{2}_{\tau}),
\end{align*}
with $\mu_{\tau}$ and $\sigma^{2}_{\tau}$ as fixed numbers. 

We use a Dirichlet process mixture prior for the subject-specific $\beta_{0i}$ parameters as
\begin{align}
\beta_{0i} &\sim N(\mu_i, \sigma^{2}_{\beta_{0i}})\, \nonumber \\ 
\mu_i|G & \sim G, \label{PHBayesMeanDPM}\\
G &\sim DP(\alpha, G_0 = N(0, \sigma^{2}_{0})). \nonumber
\end{align}

The Dirichlet process mixture prior above is essentially an infinite mixture of Normal distributions that are mixed over the mean parameter. We shall refer to this model with the Mean-DPM proportional hazards model. Alternatively, we propose a Dirichlet process mixture model that induces an infinite Normal distributions mixed overt the standard deviation parameter. This Dirichlet process prior can be written as
\begin{align}
\beta_{0i} & \sim N(0, \sigma^{2(i)}_{\beta_{0i}}), \nonumber \\ 
\sigma^{2(i)}_{\beta_{0i}}|G & \sim G,\label{PHBayesSigmaDPM} \\
G & \sim DP(\alpha, G_0 = log-Normal(\mu_{G_0}, \sigma^{2}_{G_0})). \nonumber
\end{align}
We shall refer to this new model with the above Dirichlet process mixture prior as Sigma-DPM proportional hazards model.

\section{Simulation Studies}
\label{nonCollapSim}

Using simulation studies, we investigate non-collapsibility in logistic regression and proportional hazards models. We consider three simulation scenarios: one when subject-specific intercepts are sampled independently from the standard Normal $N(\mu = 0, \sigma^{2} = 1)$, another when subject-specific intercepts are sampled from a mixture distribution of the form 
\begin{align*}
\beta_{0i} \overset{iid}{\sim} \theta_i N(\mu = -1.5, \sigma^{2} = 1) + (1 - \theta_i)N(\mu = 1.5, \sigma^{2} = 1),
\end{align*}
where $\theta_i \sim Bernoulli(p = 0.5)$ with $i \in \{1, \dots, n\}$ where $n$ is the number of subjects. Finally, in the third scenario subject-specific intercepts are sampled from a mixture distribution of the form 
\begin{align*}
\beta_{0i} \overset{iid}{\sim} \theta_i N(\mu = 0, \sigma^{2} = 1) + (1 - \theta_i)N(\mu = 0, \sigma^{2} = {5}),
\end{align*}
where $\theta_i \sim Bernoulli(p = 0.5)$ for $i \in \{1, \dots, n\}$.

We compare parameter estimation between our proposed models and some common statistical models used to analyze repeated measure binary data and survival data. For every simulation scenario, we run 1,000 simulations each with 300 subjects and 12 within-subject measurements per subject.  

\subsection{Logistic Regression Models}
Unlike linear and log links, logistic link is not collapsible. In this section, using synthetic data we compare parameter estimation under our proposed Mean-DPM and Sigma-DPM Bayesian hierarchical logistic regressions we use the following common statistical models to analyze repeated measure binary data: 

\begin{itemize}
  \item Generalized linear model with a logit link (GLM): We fit a frequentist GLM model with the logit link. This technique ignores the correlation between within-subject measurements. Further, this model does not account  for any subject-specific effect. Due to the ignorance of within subject correlations in this model, standard error for the estimated coefficients tend to underestimate the true standard error once the within subject correlation is taken into account. 
  
  \item Generalized estimating equation (GEE): Instead of a simple generalized linear model with the logit link where all within-subject measurements are treated as independent measures, one can use the generalized estimating equation framework to account for the correlation between within-subject measurements. Despite accounting for the correlation between measurements taken on the same subject, GEE does not consider any subject-specific random effect. 
  
  \item Generalized linear mixed effects model (GLMM): We also fit the frequentist generalized linear mixed effects model with subject-specific random intercepts to model binary data. GLMM is capable of taking the correlation in with-subject measurements into account. Further, GLMM is also capable of estimating subject-specific random intercepts with the assumption that the random intercepts are Normally distributed.
  
  \item Bayesian logistic regression: We also consider a Bayesian logistic regression model with a likelihood of the form
\begin{align*}  
Y_{ij} | \beta_0, \beta_1, X_{ij} &\sim Bernoulli(p_i = \beta_0 + \beta_1 X_{ij}),
\end{align*}    
where $Y_{ij}$ is the outcome of the $j^{th}$ measurement on the $i^{th}$ subject, $X_{ij}$ is the measured covariate corresponding to $Y_{ij}$ outcome, and $\beta_0$ and $\beta_1$ are intercept and slope. We assume priors of the form
\begin{align*}  
\beta_{0} &\sim N(0, \sigma^{2}_{\beta_0}), \\
\beta_{1} &\sim N(0, \sigma^{2}_{\beta_1}),
\end{align*}    
where $\sigma_{\beta_0}$ and $\sigma_{\beta_1}$ are fixed numbers. 

  \item Hierarchical Bayesian logistic regression model: Analogous to the the GLMM model to analyze binary data, one can setup a Bayesian hierarchical model with a likelihood of the form
\begin{align*}  
Y_{ij} | \beta_{0i}, \beta_0, \beta_1, X_{ij} &\sim Bernoulli(p_i = \beta_0 + \beta_1 X_{ij}),
\end{align*}    
where $Y_{ij}$ is the outcome of the $j^{th}$ measurement on the $i^{th}$ subject, $X_{ij}$ is the measured covariate corresponding to $Y_{ij}$ outcome, $\beta_{0i}$ is the subject-specific random intercepts where $i \in \{1, \dots, n\}$ with $n$ as the number of subjects in the data, and $\beta_0$ and $\beta_1$ are intercept and slope. We assume Gaussian priors on subject-specific random intercepts $\beta_{0i}$ of the form
\begin{align*}  
\beta_{0i} &\sim N(0, \sigma^{2}_{\beta_{0i}}),
\end{align*}  
where $i \in \{1, \dots, n\}$. Also, Gaussian priors are assumed on coefficients $\beta_0$, and $\beta_1$ of the form
\begin{align*}  
\beta_{0} &\sim N(0, \sigma^{2}_{\beta_0}), \\
\beta_{1} &\sim N(0, \sigma^{2}_{\beta_1}), 
\end{align*}  
where $\sigma_{\beta_0}$ and $\sigma_{\beta_1}$ are fixed numbers. 
\end{itemize}

\begin{figure}[ht!]
\centering\includegraphics[scale=0.8]{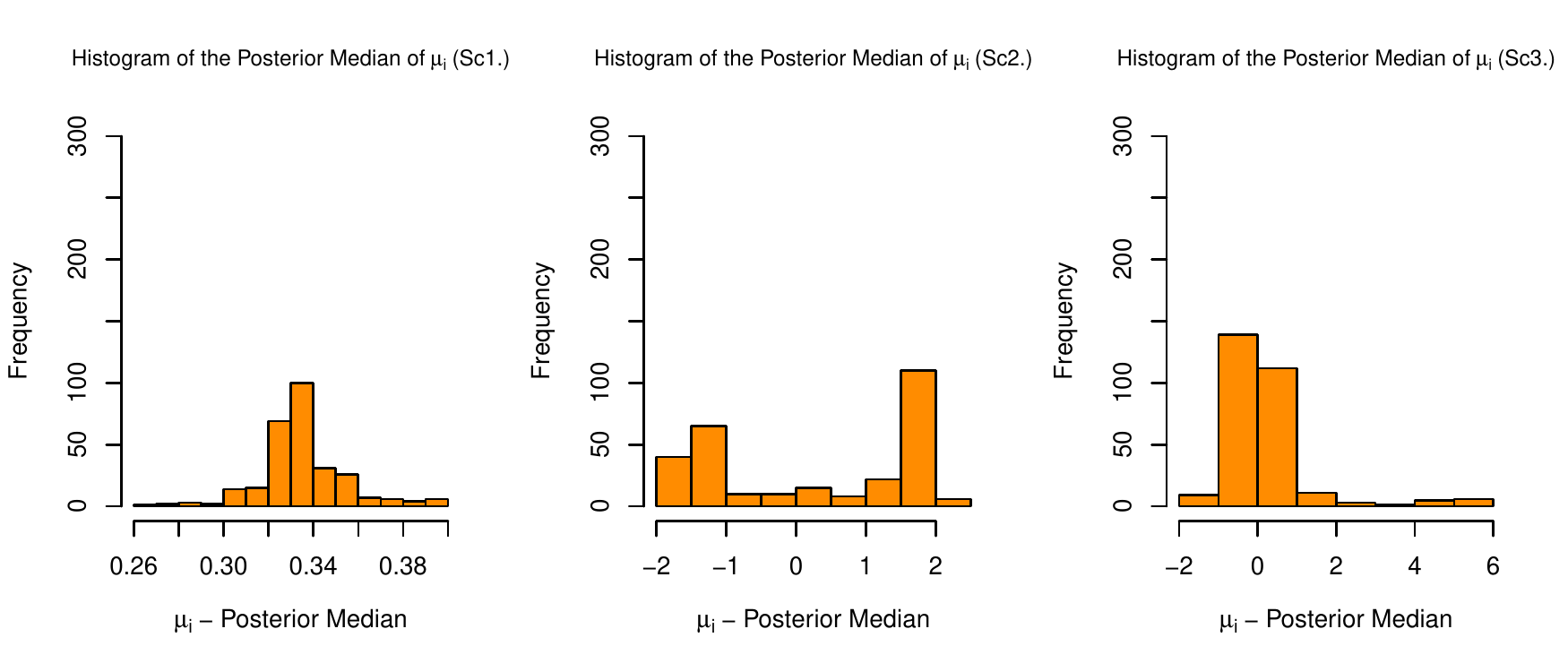}
\caption{Histogram of the posterior median of $\mu_i$'s from the proposed Mean-DPM hierarchical Bayesian logistic model, where $\mu_i$ is the subject-specific prior mean on the random intercept of subject $i$. The plot to the left is the histogram of the posterior median of the sampled $\mu_i$ from the model when it runs under the first simulation scenario where all random intercepts are sampled from the standard Normal distribution. The plot in the middle shows the histogram of the posterior medians of $\mu_i$'s under the second scenario where random intercepts are sampled from a mixture of two Normal distributions of $N(\mu_1 = -1.5, \sigma^{2} = 1)$ and $N(\mu_1 = 1.5, \sigma^{2} = 1)$ that are equally weighted. The plot to the right is the histogram of the posterior medians under the third simulation scenario where the random intercepts are simulated from the mixture of two Normal distributions of $N(\mu = 0, \sigma^{2}_{1} = 1)$ and $N(\mu = 0, \sigma^{2}_{2} = {5})$. Results, under each simulation, are from one single simulated data with $N=300$ subjects and $l_i = 12$ within subject measurements.}
\label{SimNonColLogNormMu}
\end{figure}

Figure \ref{SimNonColLogNormMu} shows the histogram of the posterior median of $\mu_i$, where $i \in \{1, \dots, n\}$ from the proposed Mean-DPM hierarchical Bayesian logistic model, where $\mu_i$ is the subject-specific prior mean on the random intercept of subject $i$ (equation (\ref{LogitBayesMeanDPM})). Under each simulation scenario, we simulated a single dataset with 300 subjects each with 12 within-subject measurements and applied our proposed Mean-DPM model. The plot to the left shows a histogram of the posterior median of $\mu_i$ when data are simulated with random intercept $\beta_{0i}$ sampled from the standard Normal $N(\mu = 0, \sigma^{2} = 1)$. As the histogram shows, most of the posterior medians are close to zero. The histogram in the model shows the distribution of the posterior median $\mu_i$ when data are simulated with random intercepts sampled from mixture of two Normal distributions of the form $\theta_i N(\mu_1 = -1.5, \sigma^{2} = 1) + (1 - \theta_i) N(\mu_1 = 1.5, \sigma^{2} = 1)$, where $\theta$ is distributed Bernoulli with parameter $p = 0.5$. As the histogram in the middle shows, posterior medians are bi-modal where modes are around the true values of -1.5 and 1.5. Finally, the histogram to the right shows the posterior median of $\mu_i$ when data are simulated with random intercepts sampled from mixture of two Normal distributions of the form $\theta_i N(\mu = 0, \sigma^{2}_1 = 1) + (1 - \theta_i) N(\mu = 0, \sigma^{2}_{2} = {5})$. Due to the differences in the standard deviations, one may expect the histogram to be spread more widely compare to the first scenario, nonetheless, posterior medians are still centered around the true mean of $0$.

\begin{figure}[ht!]
\centering\includegraphics[scale=0.8]{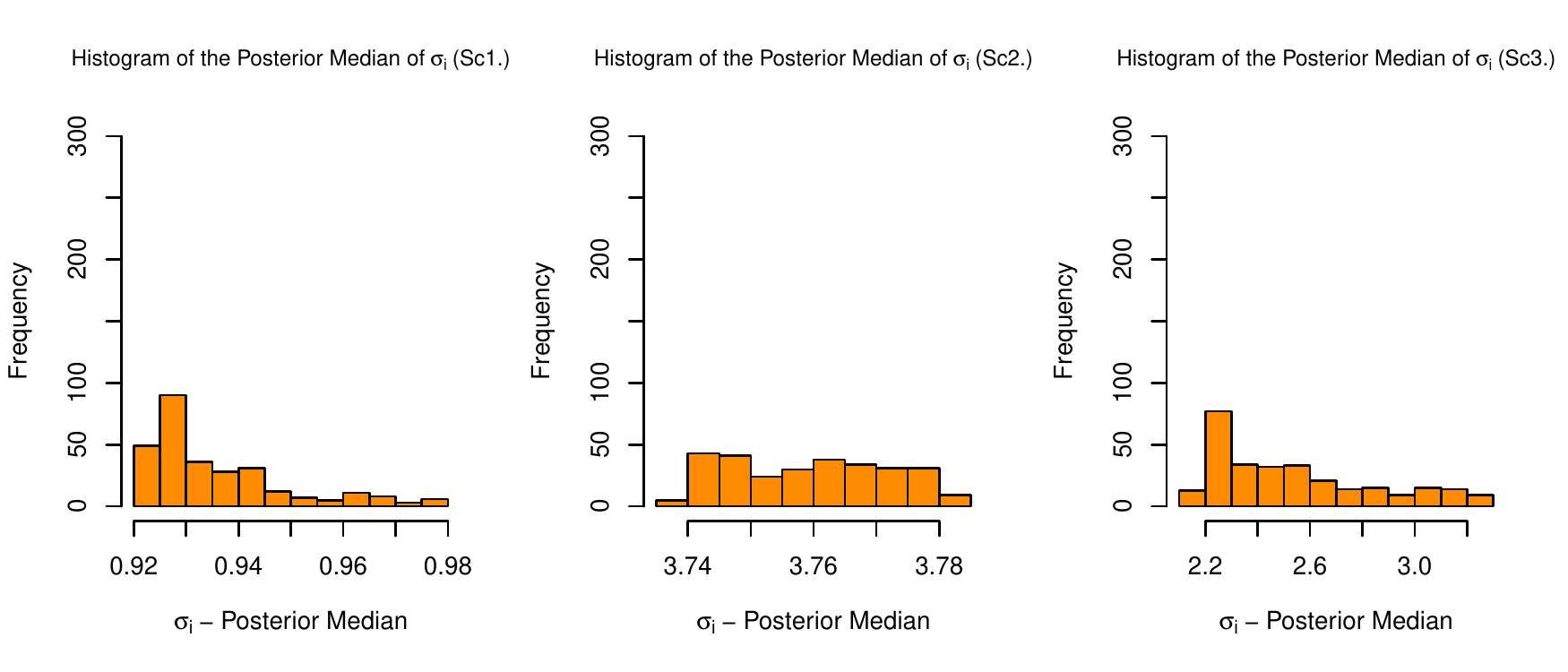}
\caption{Histogram of the posterior median of $\sigma_i$'s from the proposed Sigma-DPM hierarchical Bayesian logistic model, where $\sigma_i$ is the subject-specific prior standard deviation on the random intercept of subject $i$. The plot to the left is the histogram of the posterior median of the sampled $\sigma_i$ from the model when it runs under the first simulation scenario where all random intercepts are sampled from the standard Normal distribution. The plot in the middle shows the histogram of the posterior medians of $\sigma_i$'s under the second scenario where random intercepts are sampled from a mixture of two Normal distributions of $N(\mu_1 = -1.5, \sigma^{2} = 1)$ and $N(\mu_1 = 1.5, \sigma^{2} = 1)$ that are equally weighted. The plot to the right is the histogram of the posterior medians under the third simulation scenario where the random intercepts are simulated from the mixture of two Normal distributions of $N(\mu = 0, \sigma^{2}_{1} = 1)$ and $N(\mu = 0, \sigma^{2}_{2} = {5})$. Results, under each simulation, are from one single simulated data with $N=300$ subjects and $l_i = 12$ within subject measurements.}
\label{SimNonColLogNormSigma}
\end{figure}

Figure \ref{SimNonColLogNormSigma} shows the histogram of the posterior median of $\sigma_i$, where $i \in \{1, \dots, n\}$ from the proposed Sigma-DPM hierarchical Bayesian logistic model, where $\sigma_i$ is the subject-specific prior standard deviation on the random intercept of subject $i$ (equation (\ref{LogitBayesSigmaDPM})). Under each simulation scenario, we simulated a single dataset with 300 subjects each with 12 within-subject measurements and applied our proposed Sigma-DPM model. The plot to the left shows a histogram of the posterior median of $\sigma_i$ when data are simulated with random intercept $\beta_{0i}$ sampled from the standard Normal $N(\mu = 0, \sigma^{2} = 1)$. As the histogram shows, most the posterior medians are close to 1. The histogram in the model shows the distribution of the posterior median $\sigma_i$ when data are simulated with random intercepts sampled from mixture of two Normal distributions of the form $\theta_i N(\mu_1 = -1.5, \sigma^{2} = 1) + (1 - \theta_i) N(\mu_1 = 1.5, \sigma^{2} = 1)$, where $\theta$ is distributed Bernoulli with parameter $p = 0.5$. As the histogram in the middle shows, posterior medians are uniformly distributed from 3.18 to 3.28. This results make sense as now the data is widely spread with two distinct mean with a distance of $3$. Our Sigma-DPM model with prior mean $0$ on random intercepts has to have a larger standard deviation to provide a prior to cover all plausible subject-specific random intercepts $\beta_{0i}$. Finally, the histogram to the right shows the posterior median of $\sigma_i$ when data are simulated with random intercepts sampled from mixture of two Normal distributions of the form $\theta_i N(\mu = 0, \sigma^{2}_{1} = 1) + (1 - \theta_i) N(\mu = 0, \sigma^{2}_{2} = {5})$. It seems that in this case, the model converged to a standard deviation that is close $\sigma_2 = \sqrt{5}$. This makes sense since a when a random intercept $\beta_{0i}$ is plausible under the prior $N(0, \sigma^{2}_{1})$, it's also plausible under a prior with larger standard deviation. Hence, posterior medians converged to a large standard deviation that is plausible according to the random intercepts sampled from $N(0, \sigma^{2}_{2} = {5})$.

\begin{figure}[ht!]
\centering\includegraphics[scale=0.8]{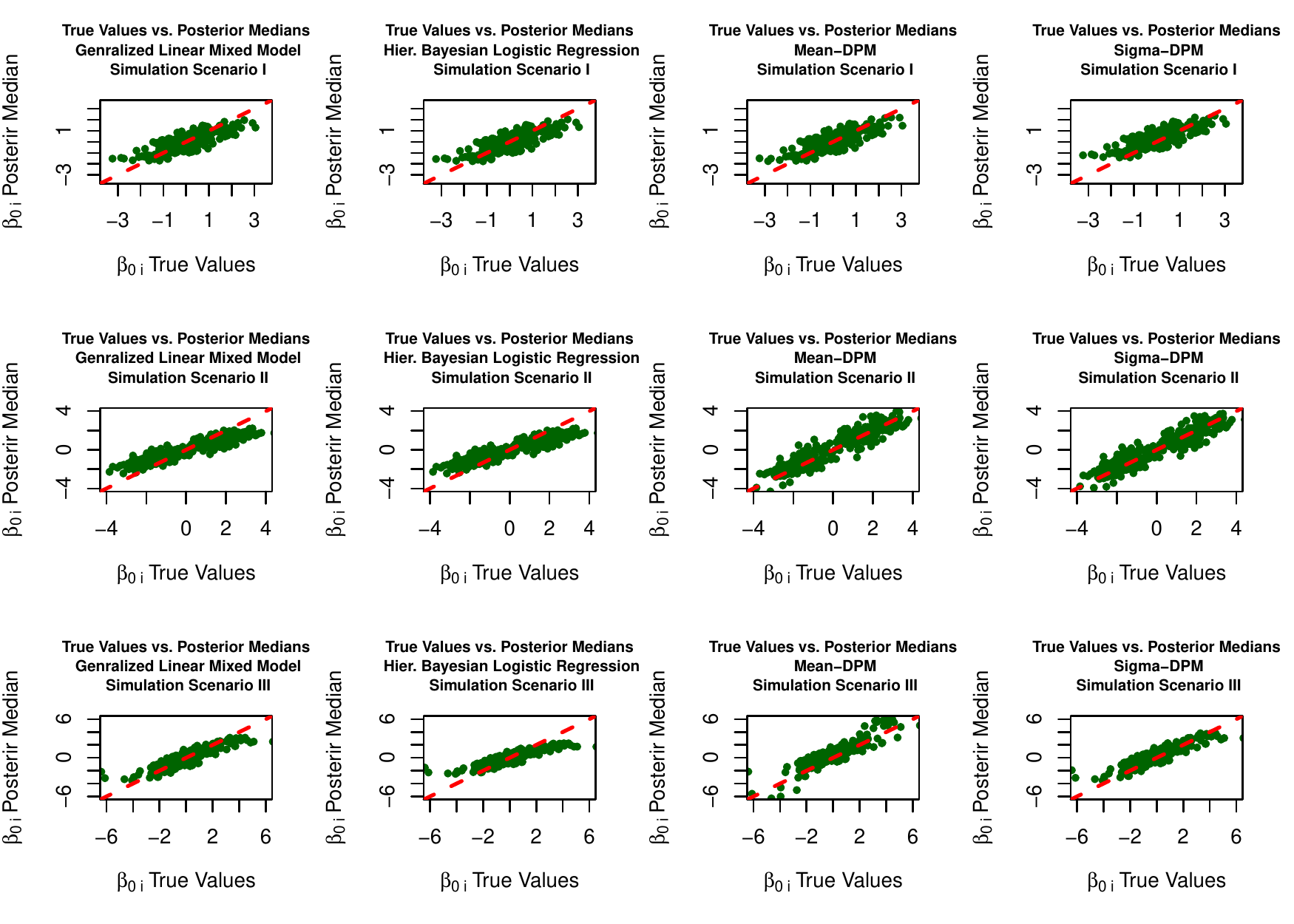}
\caption{A grid of scatter plots that shows the relation between the true values of the subject-specific random intercepts, $\beta_{0i}$, and the posterior median (or estimated) random intercepts from the GLMM model, the hierarchical Bayesian logistic model, our proposed Mean-DPM hierarchical Bayesian logistic model, and the proposed Sigma-DPM hierarchical Bayesian logistic model. The red dashed line in every plot represents the 45 degree line and the results are from a single simulated data under each simulation scenario. The first row represents the scatter plots from data simulated under the first scenario where subject-specific random intercepts are sampled from the standard Normal $N(\mu = 0, \sigma^{2} = 1)$. The second row represents scatter plots resulted from data simulated under the second simulation scenario where random intercepts are sampled from an equally weighted mixture of two Normal distributions of the form $N(\mu_1 = -1.5, \sigma^{2} = 1)$ and $N(\mu_1 = 1.5, \sigma^{2} = 1)$. Finally, the last row of plots represents results from data simulated under the third simulation scenario where random intercepts are sampled from an equally weighted mixture of two Normals of the form $N(\mu = 0, \sigma^{2}_1 = 1)$ and $N(\mu_1 = 0, \sigma^{2}_{2} = {5})$.The first column of scatter plots from left represents results from fitting the generalized linear mixed effect model, the second column represents the results from a hierarchical Bayesian logistic regression, third column represents the results from fitting our proposed Mean-DPM hierarchical Bayesian logistic model, and finally the last column to the right represents results from our proposed Sigma-DPM hierarchical Bayesian logistic model.}
\label{SimNonColLogBeta0iALL}
\end{figure}

While Figure \ref{SimNonColLogNormMu} and Figure \ref{SimNonColLogNormSigma} show the performance of our proposed models in estimating prior mean and prior standard deviation of the random intercepts, $\beta_{0i}$, however, the main interest is on evaluating the performance of the model on estimating the actual random intercepts. Figure \ref{SimNonColLogBeta0iALL} provides a grid of scatter plots each shows the relation between the true random intercept value and the posterior median or the estimate of random intercepts. As one can see in the plot, when random intercepts are Normally distributed according to the standard Normal $N(\mu = 0, \sigma^2 = 1)$ distribution, in terms of estimating the latent random intercepts, our proposed Mean-DPM and Sigma-DPM models work equally well as the GLMM model and the hierarchical Bayesian logistic model with explicit Normal assumption on the random intercepts. When the reference distribution of the sampled random intercepts is not Normal, our proposed Mean-DPM and Sigma-DPM models that are robust to distributional mis-specification of the random intercepts, outperform the GLMM and the hierarchical Bayesian logistic regression in terms of estimating the latent random intercepts. 

As tables (\ref{logistNorm}), (\ref{logistMisMean}), and (\ref{logistMisSigma}) show, coefficient estimates under marginal Bayesian model and marginal frequentist GLM and GEE shrank toward the $0$ compared to the true conditional value. The fact that in table (\ref{logistNorm}) coefficient estimates under both GLM and GEE are the same is not surprising as we are using balanced data with the canonical link. By taking sub-group intercepts into account, coefficient estimates from the generalized linear mixed effect model and the hierarchical Bayesian model with Normal prior on the random intercepts are closer to the true conditional estimand compared to the marginal models. However, the coefficient estimate under these models still shrink toward no 0. The amount of shrinkage is larger under the second and the third scenarios when the distribution of random intercepts is mis-specified. Our proposed Dirichlet process mixture models, however, are capable of detecting sub-group intercepts and are robust to distributional mis-specification of the random intercepts. Coefficient estimates from our proposed models lead to the minimum mean squared error (MSE) in estimating the true conditional coefficient value.

\begin{table}[ht]
\centering
\begin{tabular}{lccc}
  \hline
 & $\beta_x = 1.000$ & SD & MSE \\  
  \hline
GLM & 0.845 & 0.031 & 0.025 \\
GEE & 0.845 & 0.031 & 0.025 \\
Bayesian Logistic Reg. & 0.847 & 0.031 & 0.025 \\ 
GLMM & 0.951 & 0.032 & 0.004 \\ 
Hierarchical Bayes Logistic Reg. & 0.947  & 0.035 & 0.005\\
Mean-DPM Hierarchical Logistic Reg. & 1.001 & 0.036 & 0.001 \\ 
Sigma-DPM Hierarchical Logistic Reg. & 1.003 & 0.037 & 0.001 \\ 
   \hline
\end{tabular}
\caption{Binary data generated with random intercepts that are distributed according to the standard Normal distribution $N(\mu = 0, \sigma^2 = 1)$. Results are from $1,000$ different simulated data each with $N = 300$ subjects and $l_i = 12$ within subject measurements.} 
\label{logistNorm}
\end{table}

\begin{table}[ht]
\centering
\begin{tabular}{lccc}
  \hline
& $\beta_x = 1.000$ & SD & MSE \\   
  \hline
GLM & 0.626 & 0.027 & 0.141\\ 
GEE & 0.627 & 0.275 & 0.140 \\ 
Bayesian Logistic Reg. & 0.626 & 0.027 & 0.141 \\ 
GLMM & 0.938 & 0.034 & 0.005\\ 
Hierarchical Bayes Logistic Reg. & 0.931 & 0.033 & 0.006 \\ 
Mean-DPM Hierarchical Logistic Reg. & 1.006 & 0.042 & 0.002 \\
Sigma-DPM Hierarchical Logistic Reg. & 0.978 & 0.042 & 0.002 \\ 
   \hline
\end{tabular}
\caption{Binary data generated with random intercepts that are distributed according to a mixture distribution of the form $\theta_i N(\mu = -1.5, \sigma^2 = 1) + (1 - \theta_i) N(\mu = 1.5, \sigma^2 = 1)$, where $\theta_i$ are distributed $Bernoulli$ with parameter $p = 0.5$. Results are from $1,000$ different simulated data each with $N = 300$ subjects and $l_i = 12$ within subject measurements.} 
\label{logistMisMean}
\end{table}

\begin{table}[ht]
\centering
\begin{tabular}{lccc}
  \hline
& $\beta_x = 1.000$ & SD & MSE\\   
  \hline
GLM & 0.702 & 0.028 & 0.090 \\ 
GEE & 0.701 & 0.030 & 0.090  \\ 
Bayesian Logistic Reg. & 0.700 & 0.028 & 0.091\\ 
GLMM & 0.946 & 0.033 & 0.004 \\ 
Hierarchical Bayes Logistic Reg. & 0.935 & 0.034 & 0.005 \\ 
Mean-DPM Hierarchical Logistic Reg. & 0.998 & 0.041 & 0.001 \\
Sigma-DPM Hierarchical Logistic Reg. & 0.994 & 0.040 & 0.001 \\ 
   \hline
\end{tabular}
\caption{Binary data generated with random intercepts that are distributed according to a mixture distribution of the form $\theta_i N(\mu = 0, \sigma^2 = 1) + (1 - \theta_i) N(\mu = 0, \sigma^2 = 5)$, where $\theta_i$ are distributed $Bernoulli$ with parameter $p = 0.5$. Results are from $1,000$ different simulated data each with $N = 300$ subjects and $l_i = 12$ within subject measurements.} 
\label{logistMisSigma}
\end{table}

\subsection{Proportional Hazards Survival Models}\label{PHsection}
To explore non-collapsibility in proportional hazards models and to compare coefficient estimation under our proposed Mean-DPM and Sigma-DPM models with common proportional hazards models, we consider the following proportional hazards models:

\begin{itemize}
  \item The frequentist Cox model: We fit the frequentist Cox proportional hazards model. This model assumes an overall baseline hazards for all subjects. Using the partial likelihood techniques, Cox model does not need any baseline hazard specification as that measure gets canceled out during the estimation process. The Cox frequentist model does not take the differential baseline hazards across subjects into account. In fitting the Cox model, we take the within subject correlation between multiple within-subject measurements into account using the approach proposed by \cite{lee1992cox} where we first estimate model coefficients using the independent covariance matrix and then we use a robust sandwich covariance matrix to account for within subject correlation between measurements. 
  
  \item Weibull accelerated failure time model (AFT): AFT models describe survival times as a function of predictor variables. Generally, Weibull AFT models are of the form
\begin{align*}  
log(T_{ij}) = \beta_0 + \beta_1 X_{ij} + \epsilon,
\end{align*}  
where $T_{ij}$ is the survival time for the $j^{th}$ measurement on the $i^{th}$ subject, $X_{ij}$ is the corresponding covariate to the outcome $T_{ij}$, and a random error $\epsilon$ such that $T_{ij} $is distributed according to a Weibull distribution with shape parameter $\tau$ and scale parameter $exp(\lambda)$. When there exists multiple measurements per subject, failure to account for the correlation between within subject measurements leads to incorrect estimated standard error of coefficients. In order to account for this intra class correlations, we take the the approach proposed by \cite{lee1992cox} where first coefficients in the model are estimated using an independent covariance structure between within subject measurements and then a robust sandwich covariance matrix is used to account for the within cluster correlations.

  \item Bayesian marginal proportional hazards model: We consider a Bayesian proportional hazard model with a likelihood of the form
\begin{align*}  
T_{ij} | \tau, \beta_0, \beta_1, X_{ij} &\sim Weibull(\tau, \lambda_i = \beta_0 + \beta_1 X_{ij}),
\end{align*}    
where $T_{ij}$ and $X_{ij}$ are the survival times and the measured covariate on the $j^{th}$ measurement on the $i^{th}$ subject, $\tau$ is the shape parameter, $\beta_0$ and $\beta_1$ are the intercepts and the slope with $\beta_1$ as the log relative risk of death per every one unit change in $X_{ij}$. Similar to the previously introduced Weibull distribution for survival times, $\lambda_i$ is the log of the subject-specific scale parameter. 
We specify a log-Normal prior on the shape parameter $\tau$ that is of the form
\begin{align*}
\tau &\sim log-Normal(\mu_{\tau}, \sigma_{\tau}),
\end{align*}
where $\mu_{\tau}$ and $\sigma_{\tau}$ are fixed numbers. Also, $\beta_0$ and $\beta_1$ are assumed to have Gaussian priors of the form
\begin{align*}
\beta_0 &\sim N(0, \sigma^2_{\beta_0}),\\
\beta_1 &\sim N(0, \sigma^2_{\beta_1}).
\end{align*}
  \item Hierarchical Bayesian proportional hazards model: In order to account for the differential baseline hazard across subjects, one can consider a likelihood of the form
\begin{align*}  
T_{ij} | \tau, \beta_{0i}, \beta_0, \beta_1, X_{ij} &\sim Weibull(\tau, \lambda_i = \beta_{0i} + \beta_0 + \beta_1 X_{ij}),
\end{align*}    
where $\beta_{0i}$ can be considered as the subject-specific log baseline hazard. For this model, we assume similar to priors as the one specified for the ``Bayesian marginal proportional hazards model". Additionally, we assume $\beta_{0i}$, where $i\in \{1, \dots, n\}$, to have a Gaussian prior of the form:
\begin{align*}
\beta_{0i} \sim N(0, \sigma^2_{\beta_{0i}}),
\end{align*} 
where $\sigma_{\beta_{0i}}$ is a fixed number. 
\end{itemize}

\begin{figure}[ht!]
\centering\includegraphics[scale=0.8]{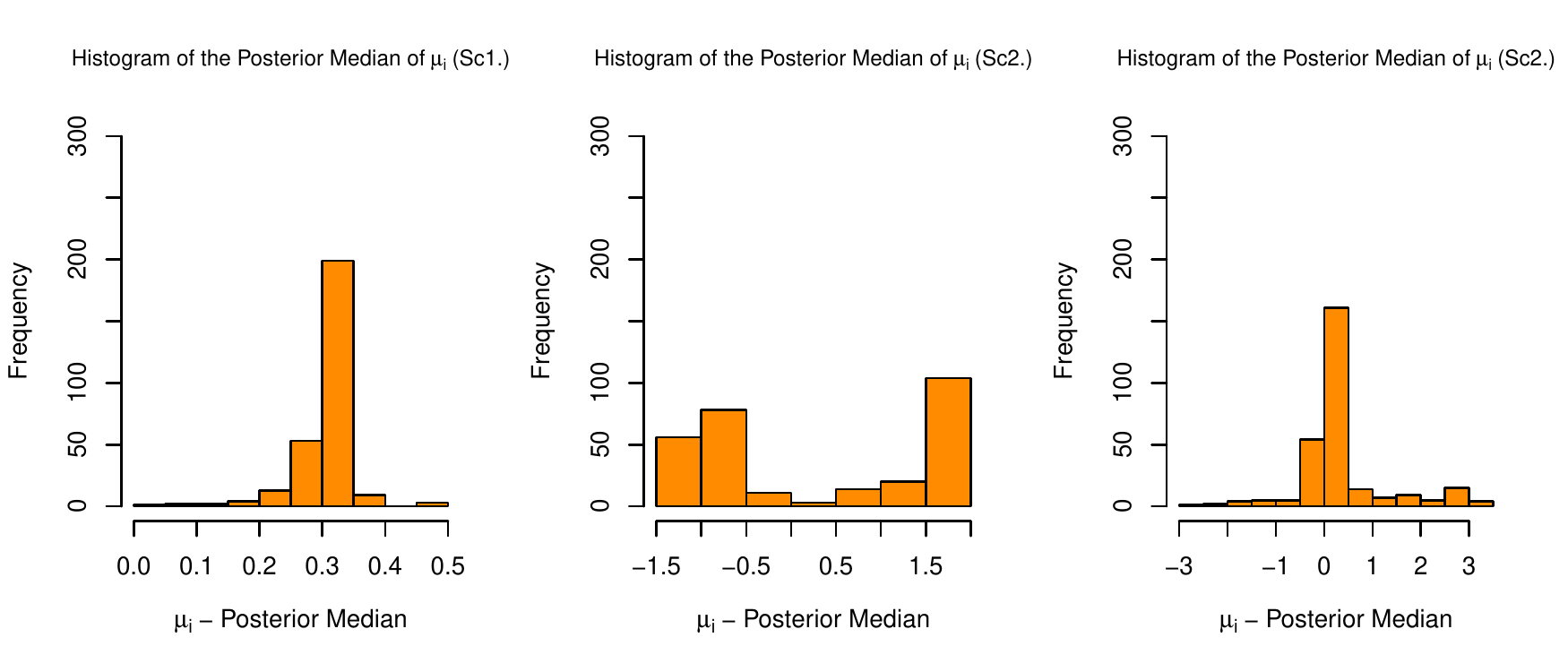}
\caption{Histogram of the posterior median of $\mu_i$'s from the proposed Mean-DPM hierarchical Bayesian proportional hazard model, where $\mu_i$ is the subject-specific prior mean on the random intercept of subject $i$. The plot to the left is the histogram of the posterior median of the sampled $\mu_i$ from the model when it runs under the first simulation scenario where all random intercepts are sampled from the standard Normal distribution. The plot in the middle shows the histogram of the posterior medians of $\mu_i$'s under the second scenario where random intercepts are sampled from a mixture of two Normal distributions of $N(\mu_1 = -1.5, \sigma = 1)$ and $N(\mu_1 = 1.5, \sigma = 1)$ that are equally weighted. The plot to the right is the histogram of the posterior medians under the third simulation scenario where the random intercepts are simulated from the mixture of two Normal distributions of $N(\mu = 0, \sigma_1 = 1)$ and $N(\mu = 0, \sigma_2 = \sqrt{5})$. Results, under each simulation, are from one single simulated data with $N=300$ subjects and $l_i = 12$ within subject measurements.}
\label{SimNonColPHNormMu}
\end{figure}

Figure \ref{SimNonColPHNormMu} shows the histogram of the posterior median of $\mu_i$, where $i \in \{1, \dots, n\}$ from the proposed Mean-DPM hierarchical Bayesian proportional hazard model, where $\mu_i$ is the subject-specific prior mean on the subject-specific log baseline hazard of subject $i$, which we represent it with $\beta_{0i}$ and for the sake consistency, we shall refer to it as the subject-specific random intercept (equation (\ref{PHBayesMeanDPM})). Under each simulation scenario, we simulated a single dataset with 300 subjects each with 12 within-subject measurements and applied our proposed Mean-DPM model. The plot to the left shows a histogram of the posterior median of $\mu_i$ when data are simulated with random intercept $\beta_{0i}$ sampled from the standard Normal $N(\mu = 0, \sigma = 1)$. As the histogram shows, most the posterior medians are close to zero. The histogram in the model shows the distribution of the posterior median $\mu_i$ when data are simulated with random intercepts sampled from mixture of two Normal distributions of the form $\theta_i N(\mu_1 = -1.5, \sigma^{2} = 1) + (1 - \theta_i) N(\mu_1 = 1.5, \sigma^{2} = 1)$, where $\theta$ is distributed Bernoulli with parameter $p = 0.5$. As the histogram in the middle shows, posterior medians are bi-modal where modes are around the true values of -1.5 and 1.5. Finally, the histogram to the right shows the posterior median of $\mu_i$ when data are simulated with random intercepts sampled from mixture of two Normal distributions of the form $\theta_i N(\mu = 0, \sigma^{2}_{1} = 1) + (1 - \theta_i) N(\mu = 0, \sigma^{2}_{2} = {5})$. Due the the differences in the standard deviations, one may expect the histogram to be spread more widely compare to the first scenario, nonetheless, posterior medians are still centered around the true mean of $0$.

\begin{figure}[ht!]
\centering\includegraphics[scale=0.8]{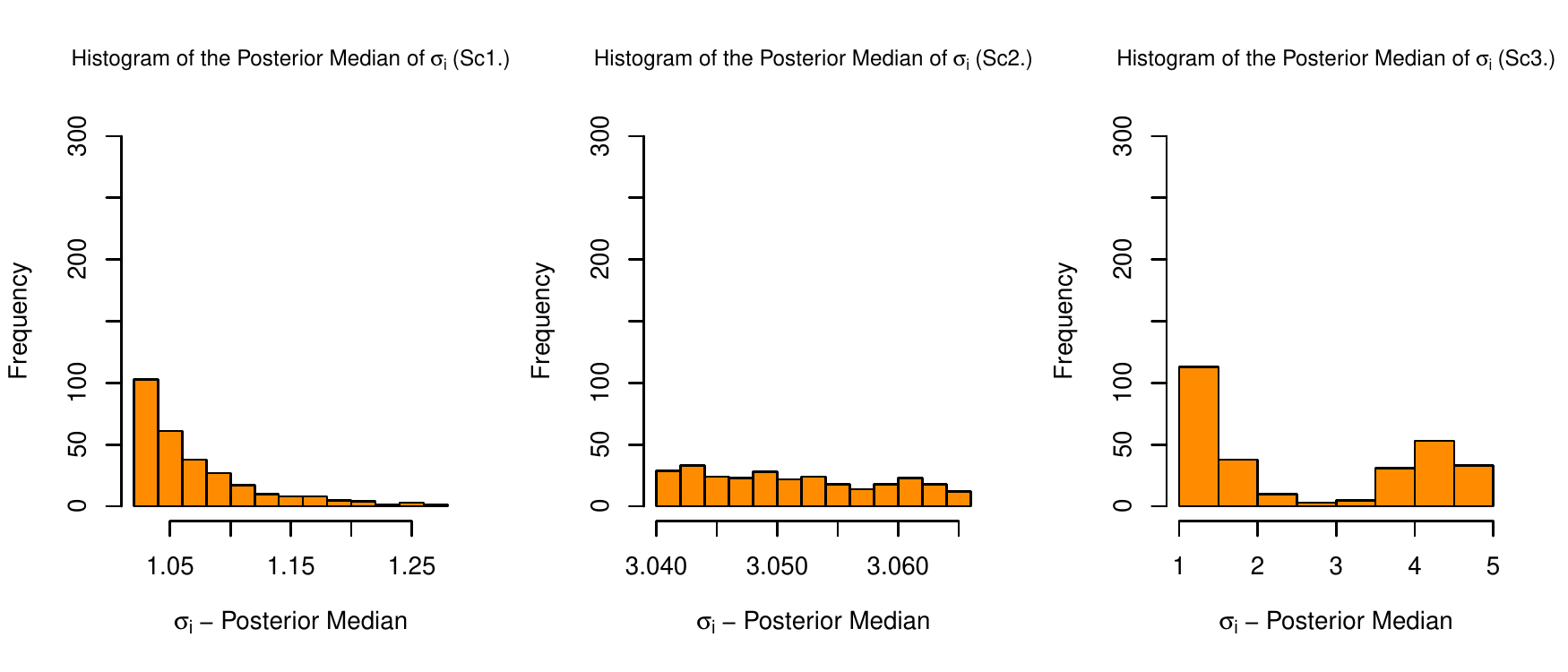}
\caption{Histogram of the posterior median of $\sigma_i$'s from the proposed Sigma-DPM hierarchical Bayesian proportional hazard model, where $\sigma_i$ is the subject-specific prior standard deviation on the random intercept of subject $i$. The plot to the left is the histogram of the posterior median of the sampled $\sigma_i$ from the model when it runs under the first simulation scenario where all random intercepts are sampled from the standard Normal distribution. The plot in the middle shows the histogram of the posterior medians of $\sigma_i$'s under the second scenario where random intercepts are sampled from a mixture of two Normal distributions of $N(\mu_1 = -1.5, \sigma = 1)$ and $N(\mu_1 = 1.5, \sigma = 1)$ that are equally weighted. The plot to the right is the histogram of the posterior medians under the third simulation scenario where the random intercepts are simulated from the mixture of two Normal distributions of $N(\mu = 0, \sigma_1 = 1)$ and $N(\mu = 0, \sigma_2 = \sqrt{5})$. Results, under each simulation, are from one single simulated data with $N=300$ subjects and $l_i = 12$ within subject measurements.}
\label{SimNonColPHNormSigma}
\end{figure}

Figure \ref{SimNonColPHNormSigma} shows the histogram of the posterior median of $\sigma_i$, where $i \in \{1, \dots, n\}$ from the proposed Sigma-DPM hierarchical Bayesian proportional hazard model, where $\sigma_i$ is the subject-specific prior standard deviation on the random intercept of subject $i$ (equation (\ref{PHBayesSigmaDPM})). Under each simulation scenario, we simulated a single dataset with 300 subjects each with 12 within-subject measurements and applied our proposed Sigma-DPM model. The plot to the left shows a histogram of the posterior median of $\sigma_i$ when data are simulated with random intercept $\beta_{0i}$ sampled from the standard Normal $N(\mu = 0, \sigma = 1)$. As the histogram shows, most the posterior medians are close to 1. The histogram in the model shows the distribution of the posterior median $\sigma_i$ when data are simulated with random intercepts sampled from mixture of two Normal distributions of the form $\theta_i N(\mu_1 = -1.5, \sigma = 1) + (1 - \theta_i) N(\mu_1 = 1.5, \sigma = 1)$, where $\theta$ is distributed Bernoulli with parameter $p = 0.5$. As the histogram in the middle shows, posterior medians are uniformly distributed from 3.18 to 3.28. This results make sense as now the data is widely spread with two distinct mean with a distance of $3$. Our Sigma-DPM model with prior mean $0$ on random intercepts has to have a larger standard deviation to provide a prior to cover all plausible subject-specific random intercepts $\beta_{0i}$. Finally, the histogram to the right shows the posterior median of $\sigma_i$ when data are simulated with random intercepts sampled from mixture of two Normal distributions of the form $\theta_i N(\mu = 0, \sigma_1 = 1) + (1 - \theta_i) N(\mu = 0, \sigma_2 = \sqrt{5})$. It seems that in this case, the model converged to a standard deviation that is close $\sigma_2 = \sqrt{5}$. This makes sense since a when a random intercept $\beta_{0i}$ is plausible under the prior $N(0, \sigma_1)$, it's also plausible under a prior with larger standard deviation. Hence, posterior medians converged to a large standard deviation that is plausible according to the random intercepts sampled from $N(0, \sigma_2 = \sqrt{5})$.

\begin{figure}[ht!]
\centering\includegraphics[scale=0.8]{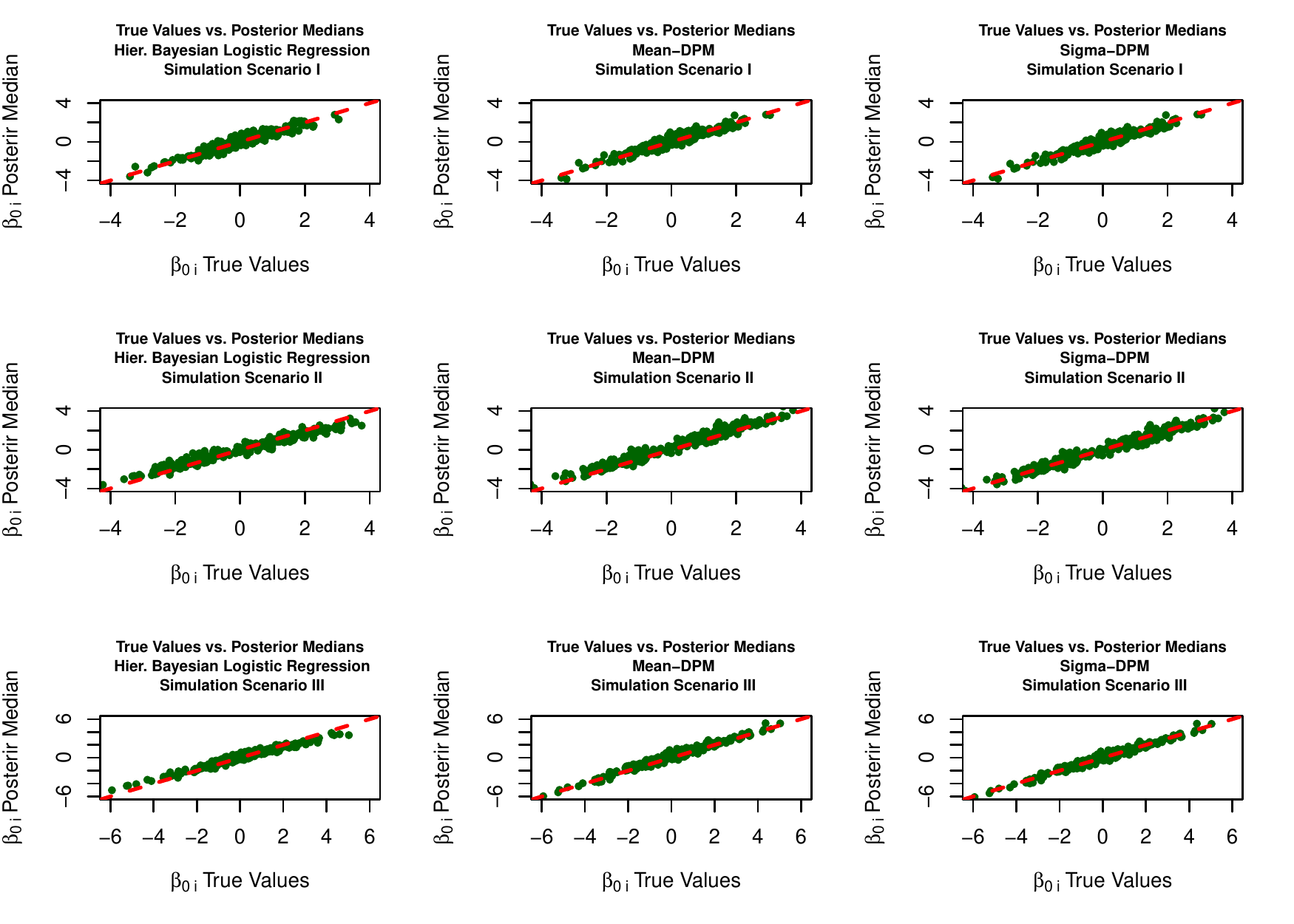}
\caption{A grid of scatter plots that shows the relation between the true values of the subject-specific random intercepts, $\beta_{0i}$, and the posterior median of random intercepts from the hierarchical Bayesian proportional hazard model, our proposed Mean-DPM hierarchical Bayesian proportional hazard model, and the proposed Sigma-DPM hierarchical Bayesian proportional hazard model. The red dashed line in every plot represents the 45 degree line and the results are from a single simulated data under each simulation scenario. The first row represents the scatter plots from data simulated under the first scenario where subject-specific random intercepts are sampled from the standard Normal $N(\mu = 0, \sigma^2 = 1)$. The second row represents scatter plots resulted from data simulated under the second simulation scenario where random intercepts are sampled from an equally weighted mixture of two Normal distributions of the form $N(\mu_1 = -1.5, \sigma^2 = 1)$ and $N(\mu_1 = 1.5, \sigma^2 = 1)$. Finally, the last row of plots represents results from data simulated under the third simulation scenario where random intercepts are sampled from an equally weighted mixture of two Normals of the form $N(\mu = 0, \sigma^2_1 = 1)$ and $N(\mu_1 = 0, \sigma^2_2 = 5)$.The first column of scatter plots from left represents results from fitting the hierarchical Bayesian proportional hazard regression, the second column represents the results from fitting our proposed Mean-DPM hierarchical Bayesian logistic model, and finally the last column to the right represents results from our proposed Sigma-DPM hierarchical Bayesian proportional hazard model.}
\label{SimNonColPHBeta0iALL}
\end{figure}

Based on Figure \ref{SimNonColPHNormMu} and Figure \ref{SimNonColPHNormSigma}, our proposed models show good performance when estimating the prior mean and prior standard deviation of the random intercepts, $\beta_{0i}$, however, the main interest is on evaluating the performance of the proposed model on estimating the actual random intercepts. In Figure \ref{SimNonColPHBeta0iALL}, we provide a grid of scatter plots each shows the relation between the true random intercept value and the posterior median estimates of those random intercepts. As Figure \ref{SimNonColPHBeta0iALL} shows, when random intercepts are distributed according to the standard Normal $N(\mu = 0, \sigma = 1)$ distribution, in terms of estimating the latent random intercepts, our proposed Mean-DPM and Sigma-DPM models work equally well as the the hierarchical Bayesian proportional hazard model with explicit Normal assumption on the random intercepts. When the reference distribution of the sampled random intercepts is not Normal, our proposed Mean-DPM and Sigma-DPM models that are robust to distributional mis-specification of the random intercepts, outperform the hierarchical Bayesian proportional hazard model in terms of estimating the latent random intercepts $\beta_{0i}$. 

Tables \ref{SurvNorm}, \ref{SurvMisMean}, and \ref{SurvMisSigma} show the results for the proportional hazards models. Coefficient estimates under the Cox model, the Bayesian marginal model, and the Weibull AFT model, all examples of marginal models, are smaller compared to the true conditional estimand and the marginal coefficient estimate under these models shrink toward $0$.

By taking the differential subject-specific baseline hazard into account, the hierarchical Bayes model with the Normal prior on random intercepts $\beta_{0i}$ is capable of estimating the true conditional estimand when the random intercepts are truly Normally distributed (Table \ref{SurvNorm}). However, the model is not robust to distributional mis-specification as under the second and the third scenarios, the coefficient estimate of $\beta_1$ shrank toward 0 (Table \ref{SurvMisMean} and Table \ref{SurvMisSigma}). 

Finally, our proposed Mean-DPM and Sigma-DPM proportional hazards models assume no explicit distributional assumption on the random intercepts, are capable of detecting subject-specific random intercepts, and are robust to distributional mis-specification of the random intercepts. Hence, our proposed DPM proportional hazard models can estimate the true conditional estimand.  

\textbf{}
\begin{table}[ht]
\centering
\begin{tabular}{lccc}
  \hline
& $\beta_x = 1.000$ & SD & MSE\\   
  \hline
Frequentist Cox Model & 0.661 & 0.089 & 0.123 \\ 
Weibull\_AFT & 0.709 & 0.096 & 0.095 \\ 
Bayesian Marginal Proportional Hazard Model & 0.700 & 0.038 & 0.100  \\ 
Hierarchical Bayesian Proportional Hazard Model & 1.015 & 0.122 & 0.014 \\ 
Mean-DPM Proportional Hazard Model & 0.995 & 0.124 & 0.015 \\ 
Sigma-DPM Proportional Hazard Model & 0.999 & 0.122 & 0.016 \\ 
   \hline
\end{tabular}
\caption{Time-to-event data generated with differential subject-specific log baseline hazards induced by subject-specific random intercepts that are distributed according to a standard Normal distribution $N(\mu = 0, \sigma^2 = 1)$. Results are from $1,000$ different simulated data each with $N = 300$ subjects and $l_i = 12$ within subject measurements.} 
\label{SurvNorm}
\end{table}

\begin{table}[ht]
\centering
\begin{tabular}{lccc}
  \hline
& $\beta_x = 1.000$ & SD & MSE\\   
  \hline
Frequentist Cox Model & 0.471 & 0.101 & 0.290 \\ 
Weibull\_AFT & 0.507 & 0.107 & 0.255  \\ 
Bayesian Marginal Proportional Hazard Model & 0.506 & 0.038 & 0.257  \\ 
Hierarchical Bayesian Proportional Hazard Model & 0.898 & 0.122 & 0.047 \\ 
Mean-DPM Proportional Hazard Model & 1.002 & 0.170 & 0.029\\ 
Sigma-DPM Proportional Hazard Model & 1.000 & 0.209 & 0.033 \\ 
   \hline
\end{tabular}
\caption{Time-to-event data generated with differential subject-specific log baseline hazards induced by subject-specific random intercepts that are distributed according to a mixture distribution of the form $\theta_i N(\mu = -1.5, \sigma^2 = 1) + (1 - \theta_i) N(\mu = 1.5, \sigma^2 = 1)$, where $\theta_i$ are distributed $Bernoulli$ with parameter $p = 0.5$. Results are from $1,000$ different simulated data each with $N = 300$ subjects and $l_i = 12$ within subject measurements.} 
\label{SurvMisMean}
\end{table}

\begin{table}[ht]
\centering
\begin{tabular}{lccc}
  \hline
& $\beta_x = 1.000$ & SD & MSE\\   
  \hline
Frequentist Cox Model & 0.460 & 0.107 & 0.303 \\ 
Weibull\_AFT & 0.481 & 0.109 & 0.292  \\ 
Bayesian Marginal Proportional Hazard Model & 0.483 & 0.038 & 0.290 \\ 
Hierarchical Bayesian Proportional Hazard Model & 0.924 & 0.121 & 0.037\\ 
Mean-DPM Proportional Hazard Model & 1.014 & 0.184 & 0.029\\ 
Sigma-DPM Proportional Hazard Model & 0.997 & 0.206 & 0.046\\ 
   \hline
\end{tabular}
\caption{Time-to-event data generated with differential subject-specific log baseline hazards induced by subject-specific random intercepts that are distributed according to a mixture distribution of the form $\theta_i N(\mu = 0, \sigma^{2} = 1) + (1 - \theta_i) N(\mu = 0, \sigma^{2} = {5})$, where $\theta_i$ are distributed $Bernoulli$ with parameter $p = 0.5$. Results are from $1,000$ different simulated data each with $N = 300$ subjects and $l_i = 12$ within subject measurements.} 
\label{SurvMisSigma}
\end{table}

\section{Sensitivity Analysis}
\label{nonCollapSensAnalysis}

Using synthetic data, we showed that our proposed Mean-DPM and Sigma-DPM are capable of estimating latent cluster-specific intercepts and are robust to distributional mis-specification. Based on the simulation results presented in Section \ref{nonCollapSim}, in terms of MSE of estimating conditional coefficients, our proposed models outperform common frequentist and Bayesian models to analyze repeated measure binary data and survival data. In this section, we are interested in testing the sensitivity of our proposed Mean-DPM and Sigma-DPM proportional hazards models with respect to the three main parameters of the number of within unit measurements ($l_i$), the difference in mean parameter $\mu_1$ and $\mu_2$ when random intercepts are simulated from the mixture of two Normal distributions of the form $N(\mu_1, \sigma^{2})$ and $N(\mu_2, \sigma^{2})$, and the ratio between the two parameters $\sigma_1$ and $\sigma_2$ when random intercepts are simulated from the mixture of two Normal distributions of the form $N(0, \sigma^{2}_{1})$ and $N(0, \sigma^{2}_{2})$. 

\subsection{Sensitivity to $l_i$}
In this section, we test the sensitivity of our proposed Mean-DPM proportional hazards and Sigma-DPM proportional hazards models with respect to the number of within subject measurements $l_i$ and under the case where the distribution of the random random intercepts is mis-specified. We generate synthetic repeated measure binary and survival data under two scenarios - one when subject-specific intercepts are sampled from a mixture distribution of the form $\beta_{0i} \overset{iid}{\sim} \theta_i N(\mu = -1.5, \sigma^{2} = 1) + (1 - \theta_i)N(\mu = 1.5, \sigma^{2} = 1)$, and another when subject-specific intercepts are sampled from a mixture distribution of the form $\beta_{0i} \overset{iid}{\sim} \theta_i N(\mu = 0, \sigma^{2} = 1) + (1 - \theta_i)N(\mu = 0, \sigma^{2} = {5})$, where $\theta_i \sim Bernoulli(p = 0.5)$ with $i \in \{1, \dots, n\}$ and $n$ as the number of subjects. By changing the number of within subject measurements $l_i$, we test the sensitivity of our proposed models. 

\begin{figure}[ht!]
\centering\includegraphics[scale=0.8]{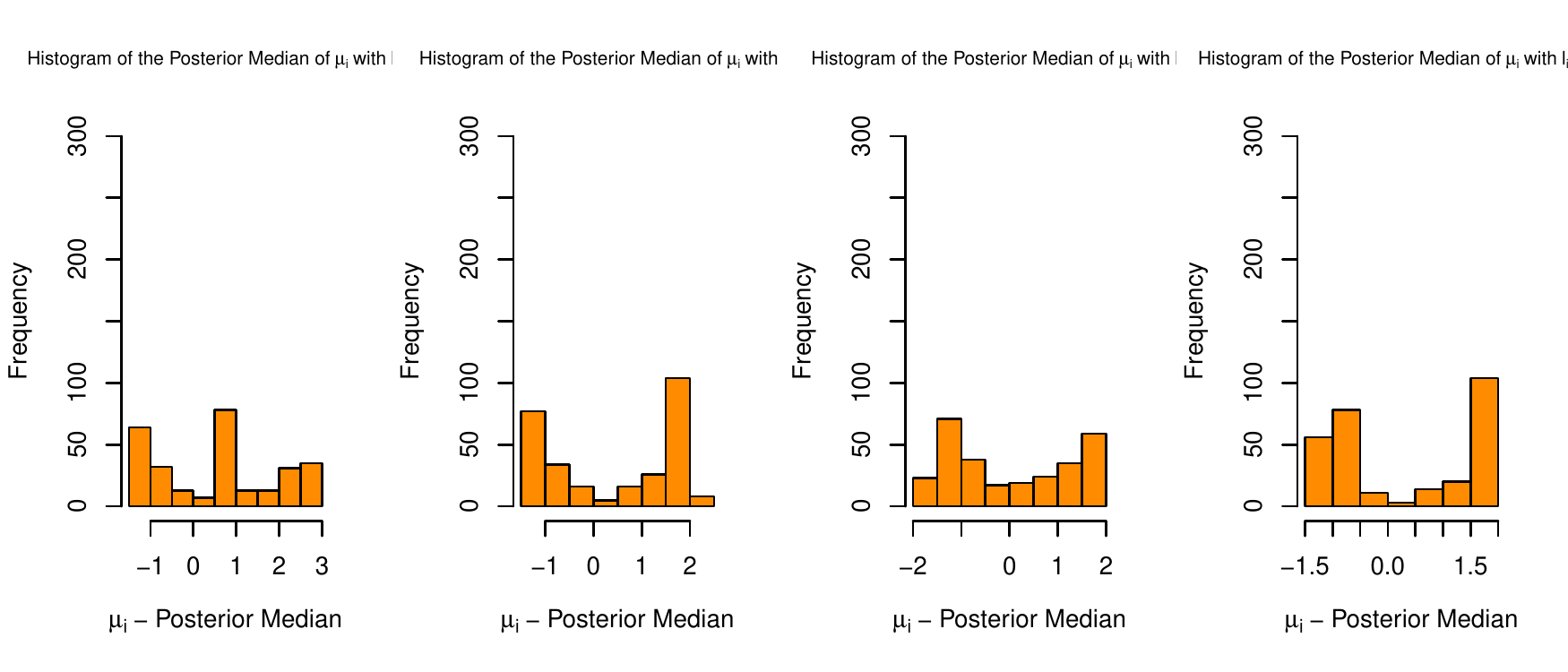}
\caption{Histogram of the posterior median of $\mu_i$'s from the proposed Mean-DPM hierarchical Bayesian proportional hazard model, where $\mu_i$ is the subject-specific prior mean on the random intercept of subject $i$. All plot are based on a simulation scenario where random intercepts are sampled from a mixture of two Normal distributions of $N(\mu_1 = -1.5, \sigma = 1)$ and $N(\mu_1 = 1.5, \sigma = 1)$ that are equally weighted. Moving from left to right, the first plots shows posterior median of $\mu_i$'s with $l_i = 1$ within subject measurement, the next plot shows the results with $l_i = 3$, the next plot shows the results under data with $l_i = 6$ within subject measurements, and finally, the last plot to the right shows the results with $l_i = 12$ within subject measurements.}
\label{SensLiMixedMuInter}
\end{figure}

Figure \ref{SensLiMixedMuInter} provides a histogram of posterior medians of the prior mean $\mu_i$ on the random intercepts $\beta_{0i}$. The results are from our proposed Mean-DPM hierarchical Bayesian proportional hazard model that is run on a single dataset that is generated under the simulation scenario where random intercepts $\beta_{0i}$'s are sampled from an equally weighted mixture of two Normal distributions with means $\mu_1 = 1.5$ or $\mu_2 = 1.5$ and with the standard deviation of $1$. As one can see, as the number of within subject measurements $l_i$ increases, our proposed Mean-DPM can better estimate the prior mean $\mu_i$'s with the true values that are either -1.5 or 1.5. 

\begin{figure}[ht!]
\centering\includegraphics[scale=0.8]{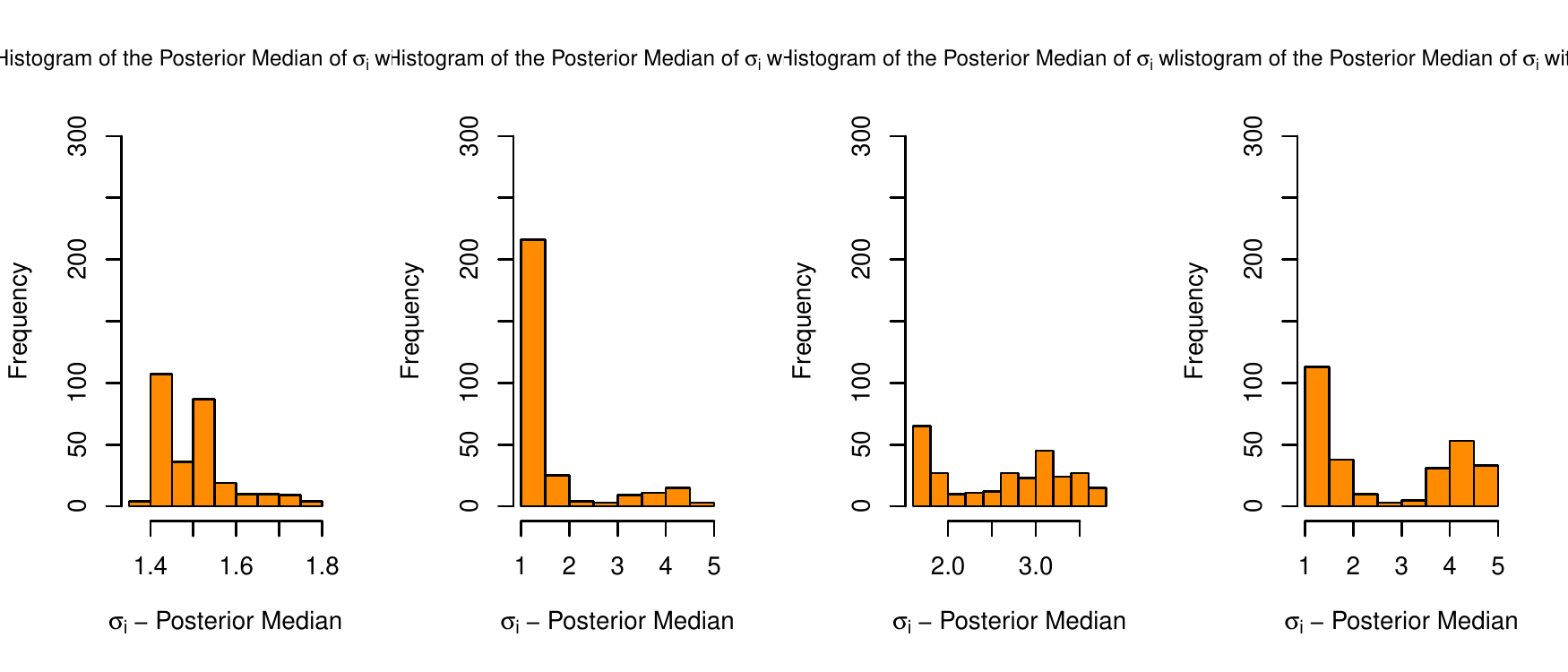}
\caption{Histogram of the posterior median of $\sigma_i$'s from the proposed Sigma-DPM hierarchical Bayesian proportional hazard model, where $\sigma_i$ is the subject-specific prior standard deviation on the random intercept of subject $i$. All plot are based on a simulation scenario where random intercepts are sampled from a mixture of two Normal distributions of $N(\mu_1 = 0, \sigma^{2} = 1)$ and $N(\mu_1 = 0, \sigma^{2} = {5})$ that are equally weighted. Moving from left to right, the first plots shows posterior median of $\sigma_i$'s with $l_i = 1$ within subject measurement, the next plot shows the results with $l_i = 3$, the next plot shows the results under data with $l_i = 6$ within subject measurements, and finally, the last plot to the right shows the results with $l_i = 12$ within subject measurements.}
\label{SensLiMixedSigmaInter}
\end{figure}

Similarly, Figure \ref{SensLiMixedSigmaInter} provides a histogram of posterior medians of the prior standard deviation $\sigma_i$ on the random intercepts $\beta_{0i}$. The results are from our proposed Sigma-DPM hierarchical Bayesian proportional hazard model that is run on a single dataset generated under the simulation scenario where random intercepts $\beta_{0i}$'s are sampled from an equally weighted mixture of two Normal distributions both with mean $\mu = 0$ and with the standard deviation of $\sigma_1 = 1$ and $\sigma_2 = \sqrt{5}$. As one can see, as the number of within subject measurements $l_i$ increases, our proposed Sigma-DPM can better estimate the prior standard deviations $\sigma_i$'s with the true values that are either $1$ or $\sqrt{5}$. 

As Figure\ref{SensLiMixedMuInter} and Figure\ref{SensLiMixedSigmaInter} show, using our proposed Mean-DPM and Sigma-DPM hierarchical Bayesian proportional hazard model, the larger within subject number of measurements, $l_i$, are, the more accurate the posterior medians of prior means $\mu_i$ and prior standard deviations $\sigma_i$ will be. $\mu_i$ and $\sigma_i$ are the hyper-parameters that are parameters of prior distributions on the random intercepts $\beta_{0i}$. 

\begin{figure}[ht!]
\centering\includegraphics[scale=0.7]{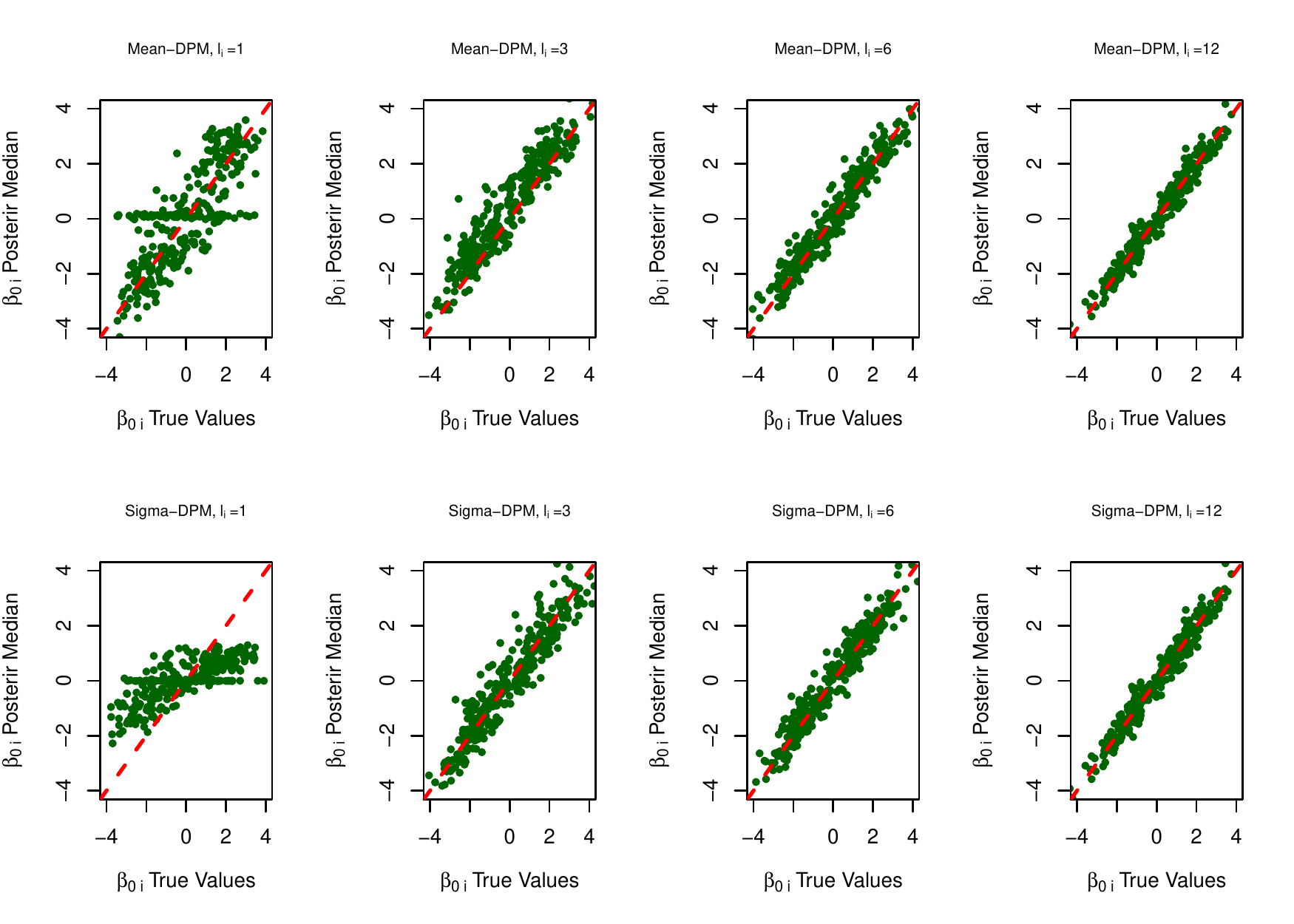}
\caption{A grid of scatter plots that shows the relation between the true values of the subject-specific random intercepts, $\beta_{0i}$, and the posterior median of random intercepts from our proposed Mean-DPM and Sigma-DPM hierarchical Bayesian proportional hazard models. The red dashed line in every plot represents the 45 degree line and the results are from a single simulated under the simulation scenario where random intercepts $\beta_{0i}$ are simulated from an equally weighted mixture of two Normal distributions one with mean $\mu_1 = -1.5$ and the other with mean $\mu_2 = 1.5$, where both distributions have the standard deviation of $\sigma = 1$. The first row represents the results from our proposed Mean-DPM and the second row represents results from our proposed Sigma-DPM model. On each row, from left to right, the scatter plots represents the results from a simulated data with $l_i = 1$, $l_i = 3$, $l_i = 6$, and $l_i = 12$ within subject measurements.}
\label{PHSensMUMIXEDInter}
\end{figure}

Figure \ref{PHSensMUMIXEDInter} includes scatterplots that show the relation between the true $\beta_{0i}$ values and the posterior medians from our proposed Mean-DPM and Sigma-DPM proportional hazard models on simulated data with the true subject-specific random intercepts $\beta_{0i}$ sampled from a mixture of two Normal distributions of the form $\theta_i N(\mu = -1.5, \sigma^{2} = 1) + (1 - \theta_i)N(\mu = 1.5, \sigma^{2} = 1)$, where $\theta_i$ is distributed Bernoulli with the parameter $p = 0.5$. As one can infer from the plots in this figure, as the number of within-subject measurements increase, posterior medians of the random intercepts provide a more accurate estimate of the true $\beta_{0i}$. 

\begin{figure}[ht!]
\centering\includegraphics[scale=0.7]{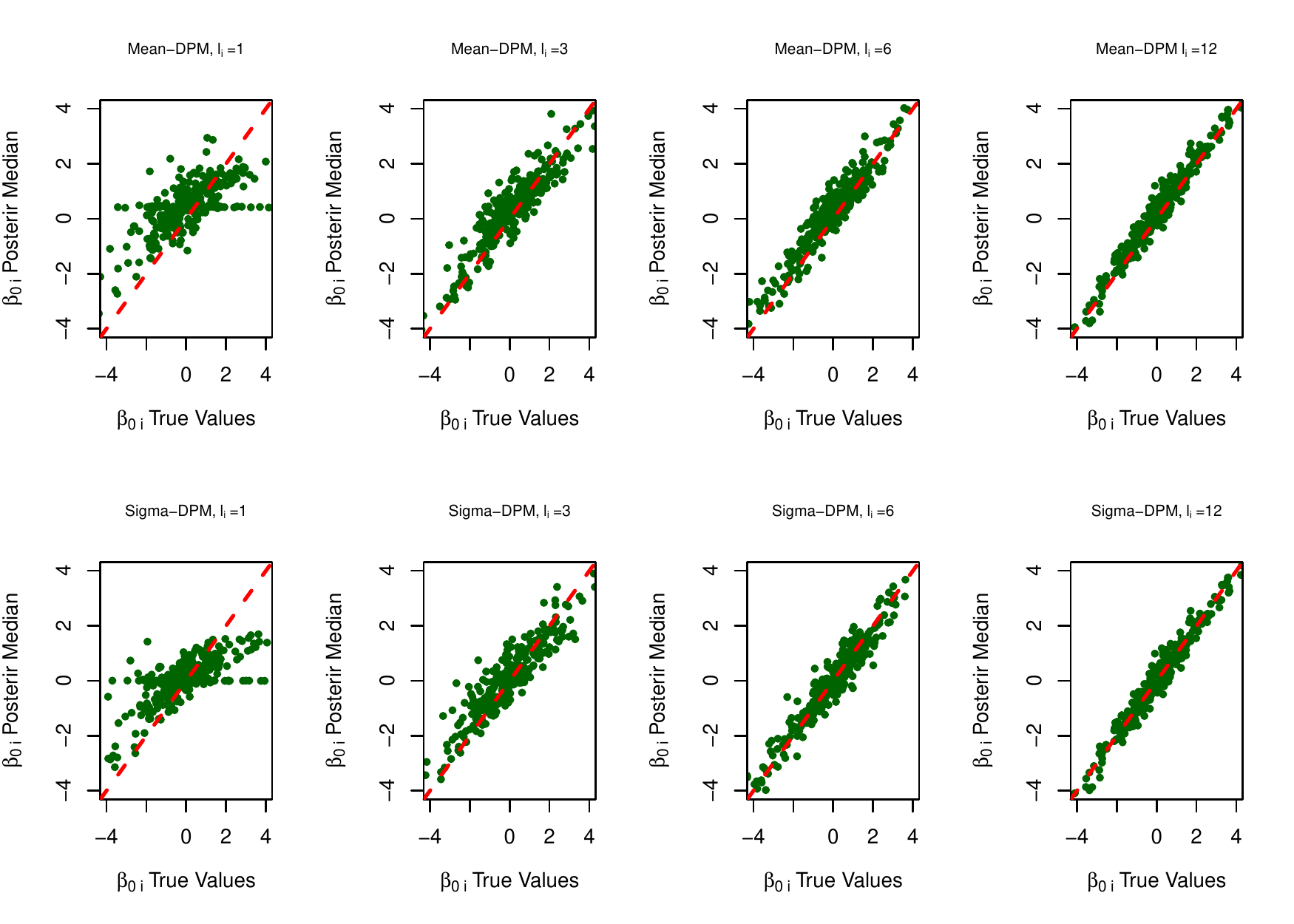}
\caption{A grid of scatter plots that shows the relation between the true values of the subject-specific random intercepts, $\beta_{0i}$, and the posterior median of random intercepts from our proposed Mean-DPM and Sigma-DPM hierarchical Bayesian proportional hazard models. The red dashed line in every plot represents the 45 degree line and the results are from a single simulated under the simulation scenario where random intercepts $\beta_{0i}$ are simulated from an equally weighted mixture of two Normal distributions both with mean $\mu = 0$ but one with the standard deviation $\sigma_1 = 1$ and another with the standard deviation of $\sigma_2 = \sqrt{5}$. The first row represents the results from our proposed Mean-DPM and the second row represents results from our proposed Sigma-DPM model. On each row, from left to right, the scatter plots represents the results from a simulated data with $l_i = 1$, $l_i = 3$, $l_i = 6$, and $l_i = 12$ within subject measurements.}
\label{PHSensSIGMAMIXEDInter}
\end{figure}

Similarly, Figure \ref{PHSensSIGMAMIXEDInter} includes similar scatterplots that show the relation between the true $\beta_{0i}$ values and the posterior medians from our proposed Mean-DPM and Sigma-DPM proportional hazard models on data simulated with the true subject-specific random intercepts $\beta_{0i}$ sampled from a mixture of two Normal distributions of the form $\theta_i N(\mu = 0, \sigma^{2} = 1) + (1 - \theta_i)N(\mu = 0, \sigma^{2} = {5})$, where $\theta_i$ is distributed Bernoulli with the parameter $p = 0.5$. From the plots in the figure, one can clearly realize that as the number of within-subject measurements increase, posterior medians of the random intercepts provide a more accurate estimate of the true $\beta_{0i}$. 

Table \ref{MeanMixtureMiSense}  provides  results on the sensitivity of our models under the first simulation scenario and table \ref{SigmaMixtureMiSense} provides the result on the sensitivity of our models under the second simulation scenario. As the results in Table \ref{MeanMixtureMiSense} and Table \ref{SigmaMixtureMiSense} show, with larger number of within subject measurements $l_i$, our proposed models can better estimate the latent random intercepts, and hence, lead to a smaller error in estimating the true conditional coefficient estimate. 

\begin{table}[ht]
\centering
\begin{tabular}{ccccccc}
  \hline
  & \multicolumn{3}{c}{Mean-DPM}  & \multicolumn{3}{c}{Sigma-DPM}  \\   
  $l_i$ & $\beta_x = 1.000$ & SD & MSE & $\beta_x = 1.000$ & SD & MSE\\   
  \hline
  1 & 0.998 & 0.223 & 0.0291 & 0.754 & 0.228 & 0.104  \\ 
  3 & 1.014 & 0.200 & 0.0283 & 0.939 & 0.218 & 0.044 \\ 
  6 & 1.010 & 0.182 & 0.033 & 0.958 & 0.211 & 0.039 \\ 
  12 & 1.002 & 0.170 & 0.029 & 1.000 & 0.209 & 0.033\\ 
   \hline
\end{tabular}
\caption{To test the sensitivity of our proposed proportional hazards models with respect to the number of within subject measurements $l_i$, time-to-event data generated with differential subject-specific log baseline hazards induced by subject-specific random intercepts that are distributed according to a mixture distribution of the form $\theta_i N(\mu = -1.5, \sigma^{2} = 1) + (1 - \theta_i) N(\mu = 1.5, \sigma^{2} = 1)$, where $\theta_i$ are distributed $Bernoulli$ with parameter $p = 0.5$. Results are from $1,000$ different simulated data each with $N = 300$ subjects and $l_i$ within subject measurements.} 
\label{MeanMixtureMiSense}
\end{table}

\begin{table}[ht]
\centering
\begin{tabular}{ccccccc}
  \hline
  & \multicolumn{3}{c}{Mean-DPM}  & \multicolumn{3}{c}{Sigma-DPM}  \\   
  $l_i$ & $\beta_x = 1.000$ & SD & MSE & $\beta_x = 1.000$ & SD & MSE\\   
  \hline
  1 & 0.939 & 0.321 & 0.077 & 0.803 & 0.237 & 0.080 \\ 
  3 & 0.984 & 0.201 & 0.031 & 0.947 & 0.201 & 0.039 \\ 
  6 & 0.987 & 0.190 & 0.039 & 0.995 & 0.210 & 0.047 \\ 
  12 & 1.014 & 0.184 & 0.046 & 0.997 & 0.206 & 0.046\\ 
   \hline
\end{tabular}
\caption{To test the sensitivity of our proposed proportional hazards models with respect to the number of within subject measurements $l_i$, time-to-event data were generated with differential subject-specific log baseline hazards induced by the subject-specific random intercept. The random intercepts are distributed according to a mixture distribution of the form $\theta_i N(\mu = 0, \sigma^{2} = 1) + (1 - \theta_i) N(\mu = 1.5, \sigma^{2} = {5})$, where $\theta_i$ are distributed $Bernoulli$ with parameter $p = 0.5$. Results are from $1,000$ different simulated data each with $N = 300$ subjects and $l_i$ within subject measurements.} 
\label{SigmaMixtureMiSense}
\end{table}

\subsection{Sensitivity to $|\mu_2 - \mu_1|$}
In this section, we test the sensitivity of our proposed Mean-DPM and Sigma-DPM proportional hazards models with respect to the distance between the mean parameters $\mu_1$ and $\mu_2$, where $\mu_1$ and $\mu_2$ are the mean parameters of two Normal distributions that are used to simulate subject-specific random intercepts. Subject-specific random intercepts are sampled from a mixture of two Normal distributions of the form $\beta_{0i} \overset{iid}{\sim} \theta_i N(\mu = -1.5, \sigma^{2} = 1) + (1 - \theta_i)N(\mu = 1.5, \sigma^{2} = 1)$, where $\theta_i$ is distributed Bernoulli with the parameter $P = 0.5$. In this section, we evaluate the sensitivity of our proposed Mean-DPM and Sigma-DPM proportional hazards models with respect to the distance between the means $\mu_1$ and $\mu_2$. In particular, we consider five cases where the distance is half of the standard deviation shared between both components, $\sigma$, or is equal to the $\sigma$, or is two times bigger than the $\sigma$, or three times bigger, or four times bigger (Table \ref{DPMmodelsMuDiffSense}). 

\begin{figure}[ht!]
\centering\includegraphics[scale=0.8]{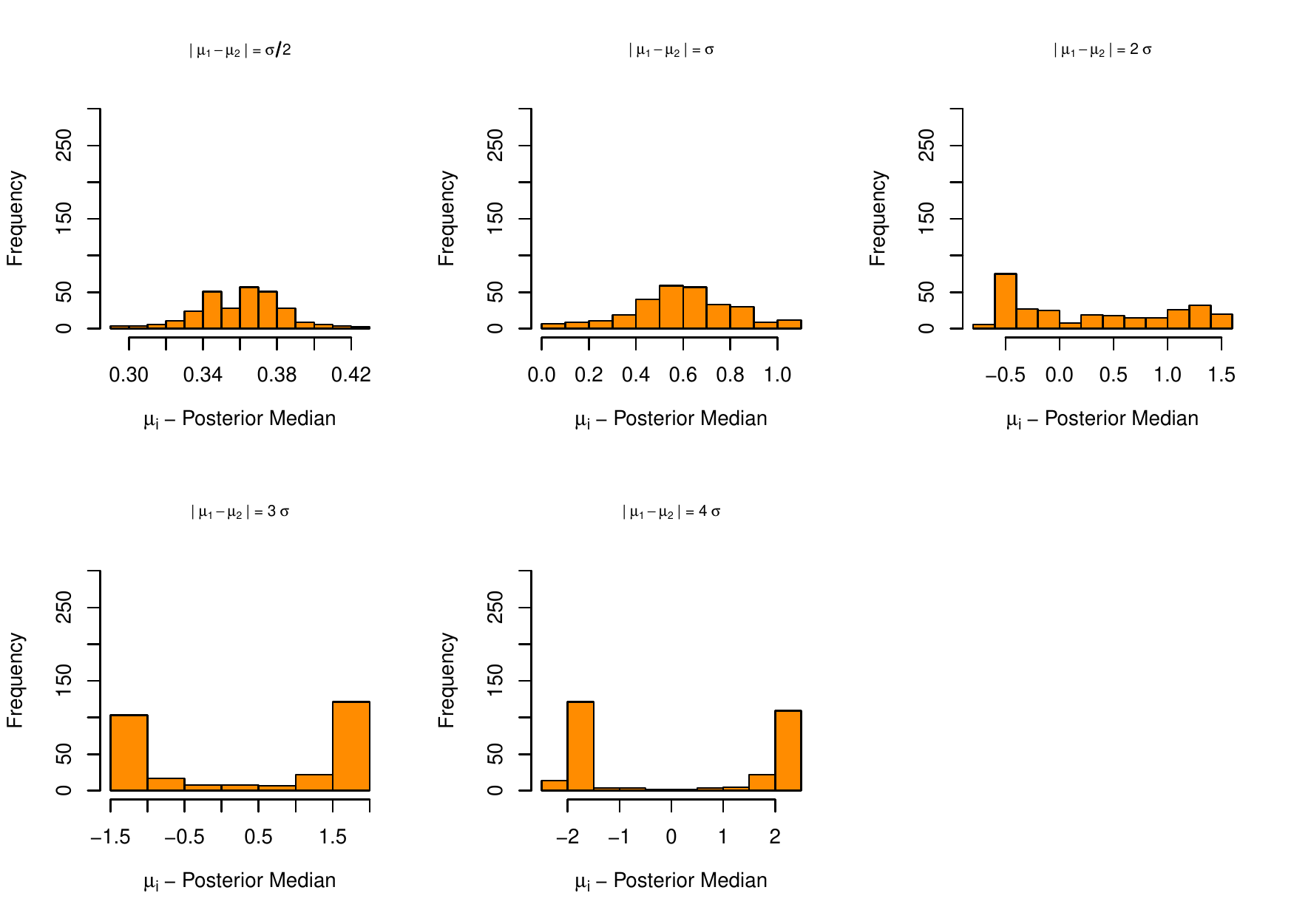}
\caption{Histogram of the posterior median of $\mu_i$'s from the proposed Mean-DPM hierarchical Bayesian proportional hazard model, where $\mu_i$ is the subject-specific prior mean on the random intercept of subject $i$. All plot are based on a simulation scenario where random intercepts are sampled from a mixture of two Normal distributions of $N(\mu_1, \sigma^{2} = 1)$ and $N(\mu_2, \sigma^{2} = 1)$ that are equally weighted with $N = 300$ subjects each with $l_i = 12$ within subject measurements. Moving from the left to right, the first plots shows posterior median of $\mu_i$'s when $\mu_1 = -0.25$ and $\mu_2 = 0.25$ (a distance of $\sigma/2$), the next plot shows the results when $\mu_1 = -0.5$ and $\mu_2 = 0.5$ (a distance of $\sigma$), the next plot is corresponding to the true $\mu_1 = -1.0$ and $\mu_2 = 1.0$ (a distance of $2\sigma$), the next plot is corresponding to the true $\mu_1 = -1.5$ and $\mu_2 = 1.5$ (a distance of $3\sigma$), the next plot is corresponding to the true $\mu_1 = -2$ and $\mu_2 = 2$ (a distance of $4\sigma$).}
\label{SensMuDiffMixedMuInter}
\end{figure}

Figure \ref{SensMuDiffMixedMuInter} provides a histogram of posterior medians of the prior mean $\mu_i$ on the random intercepts $\beta_{0i}$. The results are from our proposed Mean-DPM hierarchical Bayesian proportional hazard model that is run on a single dataset that is generated under the simulation scenario where random intercepts $\beta_{0i}$'s are sampled from an equally weighted mixture of two Normal distributions with means $\mu_1$ or $\mu_2$ and with the standard deviation of $\sigma = 1$. In order to test the sensitivity of our models with respect to the distance between $\mu_1$ and $\mu_2$, we consider 5 cases based on the distance between $\mu_1$ and $\mu_2$. Those cases are when the distance between the means is half of the standard deviation $\sigma$, equal to $\sigma$, twice of the $\sigma$, three times of the $\sigma$, or four times of the $\sigma$. 

\begin{figure}[ht!]
\centering\includegraphics[scale=0.7]{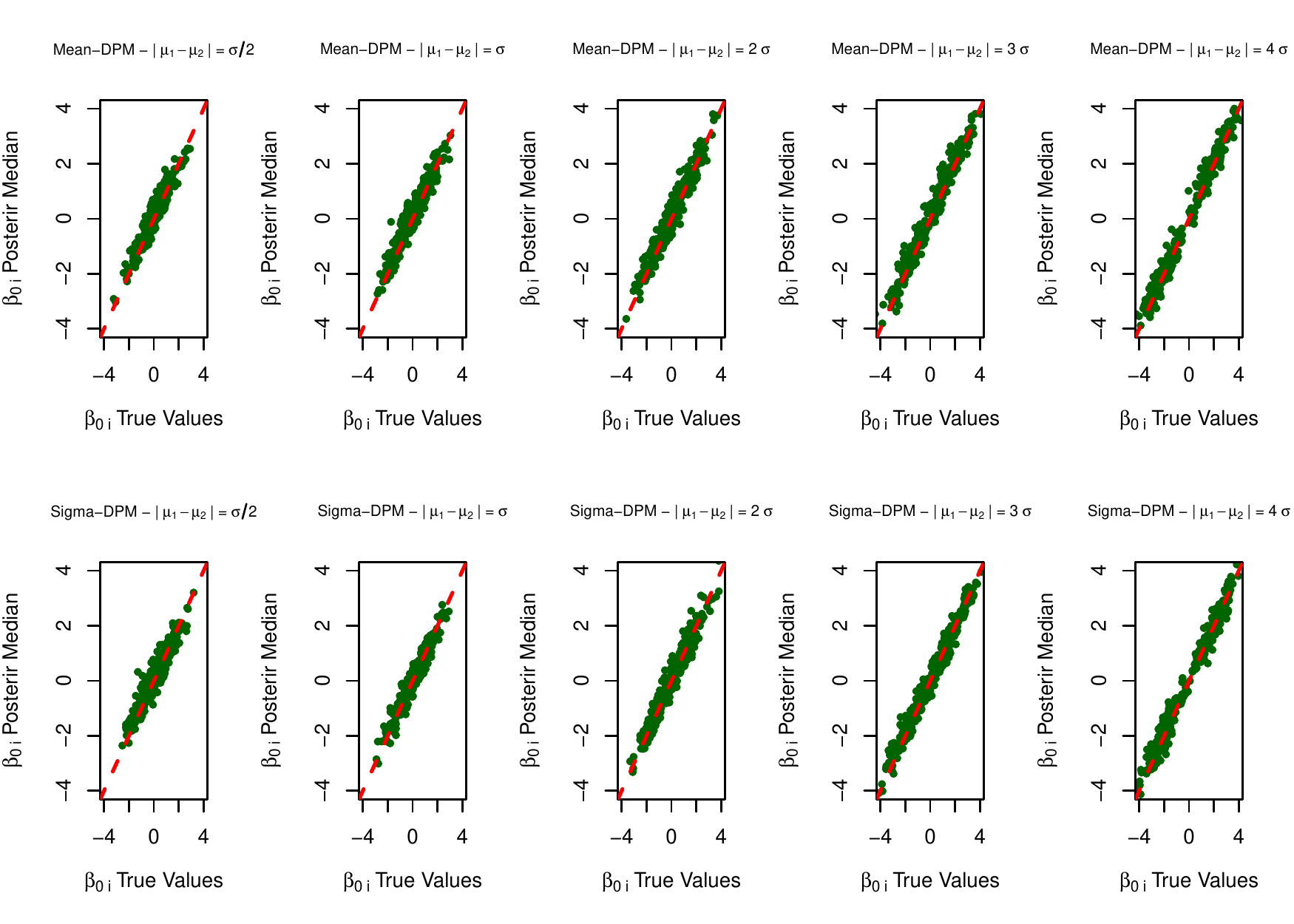}
\caption{A grid of scatter plots that shows the relation between the true values of the subject-specific random intercepts, $\beta_{0i}$, and the posterior median of random intercepts from our proposed Mean-DPM and Sigma-DPM hierarchical Bayesian proportional hazard models. The red dashed line in every plot represents the 45 degree line and the results are from a single simulated under the simulation scenario where random intercepts $\beta_{0i}$ are simulated from an equally weighted mixture of two Normal distributions of
$N(\mu_1, \sigma^{2} = 1)$ and $N(\mu_2, \sigma^{2} = 1)$. The first row represents the results from our proposed Mean-DPM and the second row represents results from our proposed Sigma-DPM model. On each row, from the left to the right, the scatter plots represents the results from a simulated data under the 5 cases of $\mu_1 = -0.25$ and $\mu_2 = 0.25$ (a distance of $\sigma/2$), $\mu_1 = -0.5$ and $\mu_2 = 0.5$ (a distance of $\sigma$), $\mu_1 = -1.0$ and $\mu_2 = 1.0$ (a distance of $2\sigma$), $\mu_1 = -1.5$ and $\mu_2 = 1.5$ (a distance of $3\sigma$), and $\mu_1 = -2$ and $\mu_2 = 2$ (a distance of $4\sigma$).}
\label{SensMuDiffAllBeta0}
\end{figure}

Figure \ref{SensMuDiffAllBeta0} includes scatterplots that show the relation between the true $\beta_{0i}$ values and the posterior medians from our proposed Mean-DPM and Sigma-DPM proportional hazard models on simulated data with the true subject-specific random intercepts $\beta_{0i}$ sampled from a mixture of two Normal distributions of the form $\theta_i N(\mu_1, \sigma^{2} = 1) + (1 - \theta_i)N(\mu_2, \sigma^{2} = 1)$,  where $\theta_i$ is distributed Bernoulli with the parameter $p = 0.5$. To test the sensitivity of our proposed models with respect to the distance between $\mu_1$ and $\mu_2$, we consider 5 different cases. Those cases are when the distance between the means are $\sigma/2$, $\sigma$, $2\sigma$, $3\sigma$, and $4\sigma$.  

As the results in Table \ref{DPMmodelsMuDiffSense} show, our proposed models are very robust in terms of the distance between the mean parameters $\mu_1$ and $\mu_2$. One may consider this fact that when $\mu_1$ and $\mu_2$ are far apart, the Dirichlet process mixture prior can easily differentiate random intercepts that are sampled from the Normal distribution with the mean $\mu_1$ from random intercepts sampled from the Normal distribution with the mean $\mu_2$. On the other hand, when $\mu_1$ and $\mu_2$ are very close, a Normal prior with an incorrectly specified mean can still cover the random intercepts that are sampled from the correct Normal distribution. Hence, our proposed models are not sensitive to the distance between the means of the Normal distributions they are sampled from.

\begin{table}[ht]
\centering
\begin{tabular}{ccccccc}
  \hline
  & \multicolumn{3}{c}{Mean-DPM}  & \multicolumn{3}{c}{Sigma-DPM}  \\   
  $|\mu_1 - \mu_2|$ & $\beta_x = 1.000$ & SD & MSE & $\beta_x = 1.000$ & SD & MSE\\   
  \hline
  $\sigma/2$ & 0.996 & 0.125 & 0.016 & 1.021 & 0.216 & 0.043 \\ 
  $\sigma$ & 1.000 & 0.132 & 0.020 & 1.019 & 0.231 & 0.049 \\ 
  $2\sigma$ & 1.021 & 0.164 & 0.027 & 0.987 & 0.285 & 0.076 \\ 
  $3\sigma$ & 1.002 & 0.170  & 0.029 & 1.000 & 0.209 & 0.033\\ 
  $4\sigma$ & 0.998 & 0.152 & 0.021 & 1.001 & 0.211 & 0.034 \\ 
   \hline
\end{tabular}
\caption{To test the sensitivity of our proposed proportional hazards models with respect to the distance between $\mu_1$ and $\mu_2$, time-to-event data were generated with differential subject-specific log baseline hazards induced by the subject-specific random intercept. The random intercepts are distributed according to a mixture distribution of the form $\theta_i N(\mu_1, \sigma^{2} = 1) + (1 - \theta_i) N(\mu_2, \sigma^{2} = 1)$, where $\theta_i$ are distributed $Bernoulli$ with parameter $p = 0.5$. Results are from $1,000$ different simulated data each with $N = 300$ subjects and $l_i = 12$ within subject measurements.} 
\label{DPMmodelsMuDiffSense}
\end{table}

\subsection{Sensitivity to $\frac{\sigma_2}{\sigma_1}$}
In this section, we test the sensitivity of our proposed Mean-DPM and Sigma-DPM proportional hazards models with respect to the relative ratio of the standard deviations $\sigma_1$ and $\sigma_2$ when the subject-specific random intercepts are sampled from a mixture of two Normal distributions of the form $\beta_{0i} \overset{iid}{\sim} \theta_i N(\mu = 0, \sigma^{2}_1) + (1 - \theta_i)N(\mu = 0, \sigma^{2}_2)$, where $\theta_i$ is distributed Bernoulli with the parameter $P = 0.5$. In this section, we evaluate the sensitivity of our proposed Mean-DPM and Sigma-DPM proportional hazards models with respect to the relative ratio of $\sigma_1$ and $\sigma_2$ that is of the from $\sigma_2/\sigma_1$. In particular, we consider four cases where the ratio 1.5, or the ratio is 2.0, or 3.0, or 5.0. As the results in Table \ref{SigmaRatioDPMSens} show, our proposed models are robust to the changes in the ratio between the standard deviations of the mixture components. 

\begin{table}[ht]
\centering
\begin{tabular}{ccccccc}
  \hline
  & \multicolumn{3}{c}{Mean-DPM}  & \multicolumn{3}{c}{Sigma-DPM}\\   
  $\sigma_2/\sigma_1$ & $\beta_x = 1.000$ & SD & MSE & $\beta_x = 1.000$ & SD & MSE\\   
  \hline
  1.5 & 0.990 & 0.255 & 0.060 & 1.001 & 0.257 & 0.069 \\ 
  2.0 & 0.988 & 0.302 & 0.100 & 1.000 & 0.304 & 0.090\\ 
  3.0 & 0.960 & 0.351 & 0.113 & 0.989 & 0.338 & 0.118\\ 
  5.0 & 1.027 & 0.368 & 0.129 & 0.976 & 0.365 & 0.129\\ 
   \hline
\end{tabular}
\caption{To test the sensitivity of our proposed proportional hazards models with respect to the ratio of $\sigma_1$ and $\sigma_2$, time-to-event data were generated with differential subject-specific log baseline hazards induced by the subject-specific random intercept. The random intercepts are distributed according to a mixture distribution of the form $\theta_i N(\mu = 0, \sigma^{2}_1) + (1 - \theta_i) N(\mu = 0, \sigma^{2}_2)$, where $\theta_i$ are distributed $Bernoulli$ with parameter $p = 0.5$. Results are from $1,000$ different simulated data each with $N = 300$ subjects and $l_i = 12$ within subject measurements.} 
\label{SigmaRatioDPMSens}
\end{table}

\begin{figure}[ht!]
\centering\includegraphics[scale=0.8]{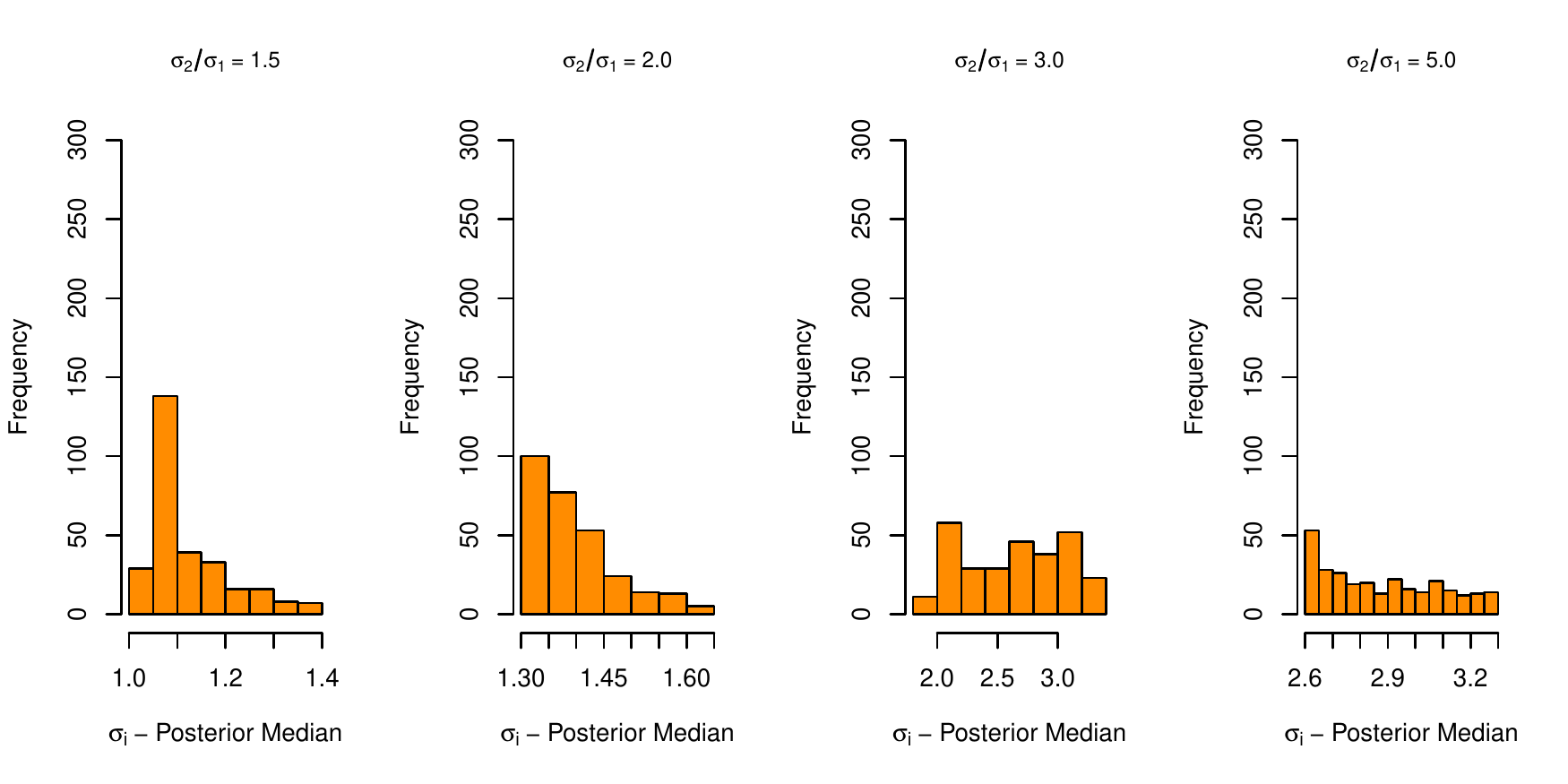}
\caption{Histogram of the posterior median of $\sigma_i$'s from the proposed Sigma-DPM hierarchical Bayesian proportional hazard model, where $\sigma_i$ is the subject-specific prior standard deviation on the random intercept of subject $i$. All plot are based on a simulation scenario where random intercepts are sampled from a mixture of two Normal distributions of $N(\mu = 0, \sigma^{2}_1)$ and $N(\mu = 0, \sigma^{2}_2)$ that are equally weighted with $N = 300$ subjects each with $l_i = 12$ within subject measurements. Moving from the left to right, the first plots shows posterior median of $\sigma_i$'s when $\sigma_1 = 1$ and $\sigma_2 = 1.5$ (a relative ratio of 1.5), the next plot shows the results when $\sigma_1 = 1$ and $\sigma_2 = 2.0$ (a relative ratio of 2.0), the next plot is corresponding to the true $\sigma_1 = 1$ and $\sigma_2 = 3.0$ (a relative ratio of 3.0), and the last plot to the right is corresponding to the true $\sigma_1 = 1.0$ and $\sigma_2 = 5.0$ (a relative ratio of 5.0).}
\label{SensSIGMADiffMixedSigmaInter}
\end{figure}

Figure \ref{SensSIGMADiffMixedSigmaInter} provides a histogram of posterior medians of the prior standard deviation $\sigma_i$ on the random intercepts $\beta_{0i}$. The results are from our proposed Sigma-DPM hierarchical Bayesian proportional hazard model that is run on a single dataset that is generated under the simulation scenario where random intercepts $\beta_{0i}$'s are sampled from an equally weighted mixture of two Normal distributions of the form $N(\mu = 0, \sigma_1)$ and $N(\mu = 0, \sigma_2)$. In order to test the sensitivity of our models with respect to the relative ratio of $\sigma_2$ and $\sigma_1$ ($\sigma_2/\sigma_1$), we consider 4 cases. Those cases are when the relative ratio of $\sigma_2/\sigma_1$ is either 1.5, or 2.0, or 3.0, or 5.0. 

\begin{figure}[ht!]
\centering\includegraphics[scale=0.7]{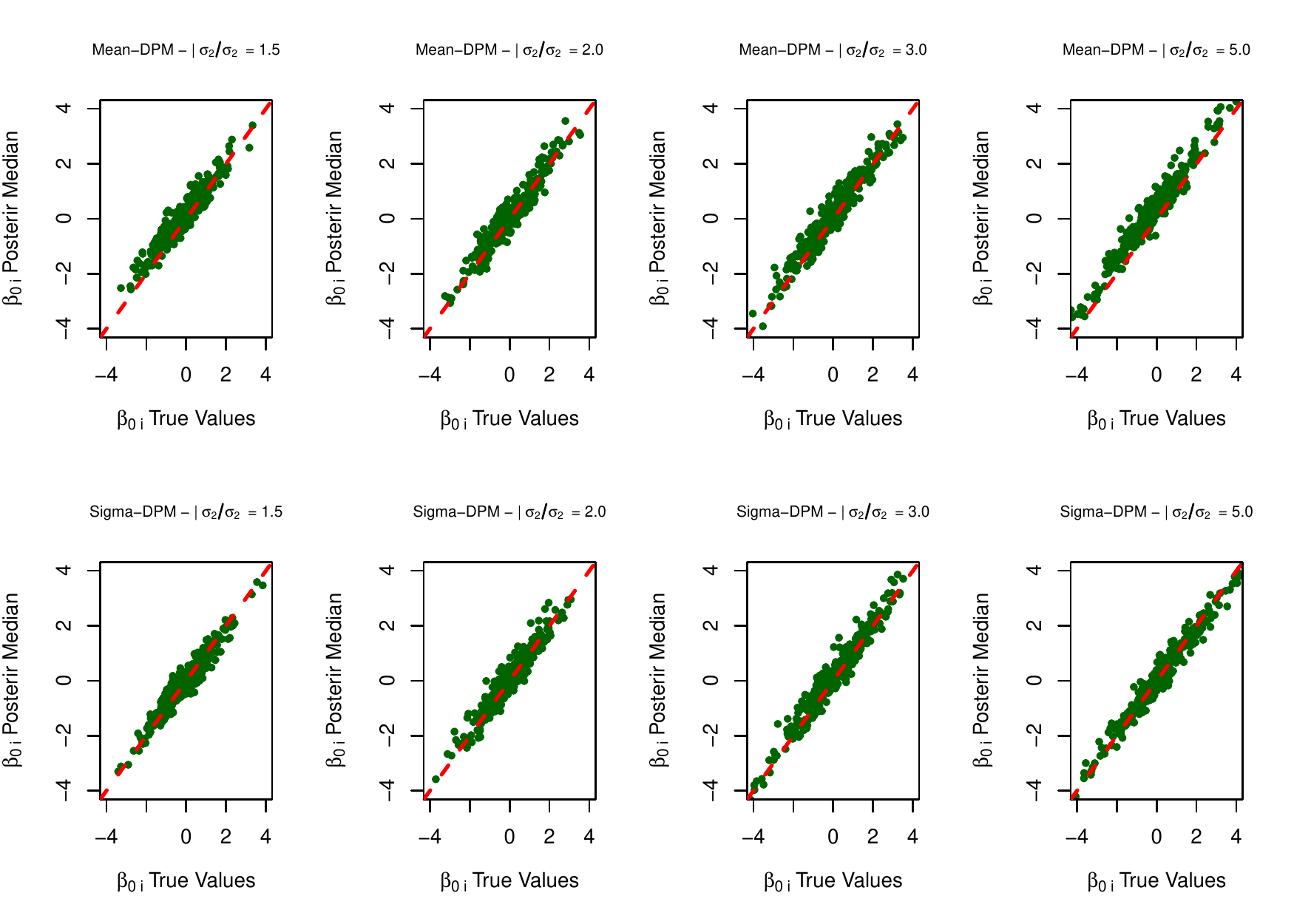}
\caption{A grid of scatter plots that shows the relation between the true values of the subject-specific random intercepts, $\beta_{0i}$, and the posterior median of random intercepts from our proposed Mean-DPM and Sigma-DPM hierarchical Bayesian proportional hazard models. The red dashed line in every plot represents the 45 degree line and the results are from a single simulated under the simulation scenario where random intercepts $\beta_{0i}$ are simulated from an equally weighted mixture of two Normal distributions of
$N(\mu = 0, \sigma^2_1)$ and $N(\mu = 0, \sigma^2_2)$. The first row represents the results from our proposed Mean-DPM and the second row represents results from our proposed Sigma-DPM model. On each row, from the left to the right, the scatter plots represents the results from a simulated data under the 4 cases of $\sigma_1 = 1.0$ and $\sigma_2 = 1.5$ (a relative ratio of 1.5), $\sigma_1 = 1.0$ and $\sigma_2 = 2.0$ (a relative ratio of 2.0), $\sigma_1 = 1.0$ and $\sigma_2 = 3.0$ (a relative ratio of 3.0), and $\sigma_1 = 1.0$ and $\sigma_2 = 5.0$ (a relative ratio of 5.0).}
\label{SensSigmaDiffAllBeta0}
\end{figure}

Figure \ref{SensSigmaDiffAllBeta0} includes scatterplots that show the relation between the true $\beta_{0i}$ values and the posterior medians from our proposed Mean-DPM and Sigma-DPM proportional hazard models under the four simulation scenarios and on simulated data with the true subject-specific random intercepts $\beta_{0i}$ sampled from a mixture of two Normal distributions of the form $\theta_i N(\mu = 0, \sigma^{2}_1) + (1 - \theta_i)N(\mu = 0, \sigma^{2}_2)$,  where $\theta_i$ is distributed Bernoulli with the parameter $p = 0.5$   

\section{Examining Different Dialysis Access Types Among Hemodialysis Patients}
\label{nonCollapApp}

End stage renal disease (ESRD) is a condition where kidneys are not capable of filtering blood from toxins. Standard care for ESRD patients are either kidney transplantation or hemodialysis. Hemodialysis is a technique that removes blood from the body through access needles and cleans the blood out of toxins using a dialysis machine. 

ESRD patients who are treated with hemodialysis, typically undergo this treatment three to four sessions a week each session three to four hours. Given the frequency of the treatment, it is unfeasible to insert a new access at every treatment session as repeatedly inserting a new access may result in irreparable damage to the patient's vein. As an alternative to a temporary access, a permanent access may be surgically placed in patient's body. Permanent accesses are in two main types of prosthetic graft and autogenous arteriovenous fistula (AVF). Prosthetic graft can be easily placed in patient's body. Similarly, AVF access can be placed as a standard attachment to a vein. When veins are hard to find, which is common among diabetic patients, AVF access is placed in the patient's body using a venous transplantation. 

Although permanent access technology has improved over time, yet access failure has remained a major issue among the hemodialysis patients. It's of interest to compare durability of different access types among hemodialysis patients. To do so, observational data were collected on 1,255 hemodialysis patients from clinics across the United States. Patients were asked to participate in the study at the time that they had their first permanent access placement. They were then followed over time prospectively and the time to failure from the time of access placement was recorded. Since patients must always have an access in order to do hemodialysis, if an access fails, the access is replaced with another access which may be of a different type than the previous access. Our data include an overall of 1,647 access records from an overall of 1,255 subjects. Some subjects may have multiple access failures during the study. In particular, over the study followup, 76.7\% of subjects had no access failure, 18\% of subjects had one access failure, 4\% had two access failures, and 1.3\% had three or more access failures.

Table \ref{AppRsltCh3} shows the result of analyzing the association between the access type and time to failure of the access. We started by analyzing the data using the Cox proportional hazards model. Note that in the case of multiple access failures per subject in the data, within subject measurements are correlated. In this case, the within subject correlation should be taken into account and standard errors of the estimated coefficients should be taken into account. To do so, we considered the approach proposed by \cite{lee1992cox} in which first the coefficients in the Cox model are estimated using maximizing the partial likelihood under an independent covariance assumption and then a robust sandwich covariance matrix is used to account for within-cluster correlations. This method is available in R programming language using the 'cluster()' function that is used inside the 'coxph()' function in order to fit a Cox model.

Next, we analyzed the data with our proposed Mean-DPM and Sigma-DPM proportional hazards model. While Cox model is not capable of taking the latent population subgroups into account and hence, coefficient estimates from this model are marginalized over all population subgroups, our proposed Mean-DPM and Sigma-DPM models, however, by accounting for the differential subject-specific baseline hazards, are capable of estimating the conditional coefficient estimates that are conditioned on subject-specific baseline hazards.

Our proposed Mean-DPM and Sigma-DPM models suggest that different access types, after adjusting for other potential risk factors in the model, are different in terms of the risk of failure. In particular, compared to the graft access, both venous and the standard fistula method have higher risk of failure. The fitted Cox model, however, finds the standard fistula access to have lower risk of failure compared to the graft method. This difference might be an indication of the attenuation in the marginal coefficient estimates under the Cox model, compared to the conditional coefficient estimate suggested by our proposed Mean-DPM model.

\begin{table}[t!]
\begin{center}
\vspace{12pt}
\scriptsize  
\begin{tabular}{lrrccccc}
\hline
& & & \multicolumn{1}{c}{Cox Model} & & \multicolumn{1}{c}{Mean-DPM Model} & \multicolumn{1}{c}{Sigma-DPM Model} \\
\cline{4-4} \cline{6-7} & No. of& No. of& \multicolumn{1}{c}{Relative Risk} & & \multicolumn{1}{c}{Relative Risk} & \multicolumn{1}{c}{Relative Risk} \\
Covariates & Cases & Failure & (95\% CI) & P-value & (95\% CR) & (95\% CR) \\
\hline
Access Type & & & & & & & \\ 
\hspace{0.5cm} graft &  1,140 & 271 & 1.0 & & 1.0 & 1.0 & \\ 
\hspace{0.5cm} standard fistula &  367 & 81&0.91 (0.70,1.18) & 0.471  & 1.11 (0.80,2.13) & 1.09 (0.73, 2.18) & \\
\hspace{0.5cm} venous transposition fistula & 140 & 40 & 1.43 (1.02,2.00) & 0.039 & 1.46 (0.96,2.19) & 1.45 (0.93,2.17) & \\
\\
Age & 1,647 & 392 & 1.00 (0.99-1.01)  & 0.513 & 1.00 (0.94,1.01) & 1.01 (0.92,1.04) & \\ 
\\
Female & 1,647 & 392 & 1.11 (0.90-1.38) & 0.328 & 1.12 (0.85,1.52) & 1.14 (0.82,1.57) & \\ 
\\
Race & & & & & & & \\ 
\hspace{0.5cm} Caucasian & 987 & 218 & 1.0 & & 1.0 & 1.0 & \\ 
\hspace{0.5cm} African American & 550 & 152 & 1.22 (0.98,1.52) & 0.071 & 1.24 (0.71,1.65) & 1.24 (0.70,1.65) & \\
\hspace{0.5cm} other & 110 & 22 & 0.79 (0.50,1.22) & 0.288 & 0.81 (0.43,1.40) & 0.82 (0.39,1.38) & \\
\\
BMI & 1,647 & 392 & 0.99 (0.98-1.01) & 0.287 & 0.99 (0.93,1.01) & 0.98 (0.91,1.03) &  \\ 
\\
Smoking & & & & & & & \\ 
\hspace{0.5cm} never smoked & 900 & 219 & 1.0 & & 1.0 & 1.0 & \\ 
\hspace{0.5cm} former smoker & 517 & 116 & 0.98 (0.78,1.24) & 0.89 & 0.98 (0.66,1.30) & 0.99 (0.64,1.32) & \\
\hspace{0.5cm} current smoker & 230 & 57 & 1.10 (0.81,1.50) & 0.541 & 1.08 (0.44,1.61) & 1.07 (0.42,1.65) & \\
\\
Serum Calcium (mg/dL) & 1,647 & 392 & 0.97 (0.87-1.08) & 0.595 & 0.99 (0.19,1.10) & 0.99 (0.18,1.13) & \\
\\
Serum Phosphorus (mg/dL) & 1,647 & 392 & 1.02 (0.96-1.07) & 0.524 & 1.00 (0.71,1.09) & 1.00 (0.68,1.10) & \\
\\
Hematocrit (g/dL) & 1,647 & 392 & 0.99 (0.97-1.01) & 0.317 & 0.99 (0.91,1.01) & 0.99 (0.90,1.04) & \\
\\
Serum Albumin (g/dL) & 1,647 & 392 & 1.01 (0.84-1.21) & 0.909 & 0.88 (0.36,1.24) & 0.83 (0.36,1.29) & \\
\\
Diabetes & 1,647 & 392 & 1.24 (1.01-1.54) & 0.041 & 1.18 (0.5,1.58) & 1.16 (0.52 ,1.51) & \\
\hline
\end{tabular}
\caption{In order to compare durability of different hemodialysis access types, observational data on 1,255 hemodialysis patients were analyzed using the Cox proportional hazards model, our proposed Mean-DPM proportional hazards model, and our proposed Sigma-DPM hazards model.}
\label{AppRsltCh3}
\end{center}
\end{table}
\normalsize

\section{Discussion}\label{nonCollapDisc}
A model with different marginal and conditional coefficient estimands is a non-collapsible model. Examples of such models include the logistic regression and the proportional hazard  models. In this paper and in the context of analyzing repeated measure data, we have proposed hierarchical Bayesian models with the Dirichlet process mixture priors. We have shown that our proposed models are capable of detecting latent subgroup effects and hence, are capable to estimate the true conditional parameters where a population consists of sub-populations with latent sub-population effects. In particular, we considered hierarchical Bayesian logistic regression and hierarchical Bayesian proportional hazards models with the Dirichlet process mixture prior on latent subgroup intercepts. We compared coefficient estimates under our proposed models with the coefficient estimates under common logistic regression and proportional hazards models. Further, we have shown that our proposed models are robust to distributional mis-specification of the latent subgroup effects. Finally, the sensitivity of our proposed models were tested in terms of their sensitivity to the number of within-cluster measurements as well as the distribution parameters of the latent cluster-specific intercepts. 

Using simulation studies, we compared coefficient estimation under our proposed Dirichlet process mixture models with common statistical longitudinal models. In particular, we compared our proposed Dirichlet process logistic regression models with the generalized linear model with the logit link, the generalized estimating equation with the logit link, the generalized linear mixed effects model with the logit link, Bayesian logistic regression, and Bayesian hierarchical logistic regression. We also compared our proposed proportional hazards models with the frequentist Cox model, the Weibull accelerated failure time model, a marginal Bayesian proportional hazards model, and a hierarchical Bayesian model. We learned that among all these models, our proposed Dirichlet process mixture models lead to the minimum mean squared errors in estimating the conditional coefficient estimands. Furthermore, while other candidate models may depend on explicit distributional assumptions over the latent sub-group random intercepts, our proposed Dirichlet process mixture models are robust to distributional mis-specification. Using sensitivity analysis, we showed that our proposed Dirichlet process mixture models are robust in terms of the number of within-cluster measurements. We have also shown that when cluster-specific random intercepts simulated from a mixture of two normal distributions, our proposed models are robust regardless of the distributional overlap of the mixing components. More generally and with the support of the simulation studies presented in this paper, in analyses aiming to characterize conditional effect of covariates using the proportional hazards or the logistic models, our proposed Dirichlet process mixture models will serve the best in terms of mean square error of estimating conditional estimands compared to other candidate models that were considered in this paper. 

Despite the capability of our proposed methods in estimating conditional estimands in repeated measure data with latent sub-group random intercepts, our proposed methods, however, are computationally demanding. Our proposed Dirichlet mixture models, on average, and for a dataset with 300 subjects each with 12 within subject measurements and using, takes 3 hours to fit using a 2.53 GHz intel Core 2 Duo processor and 4 GB 1067 MHz DDR3 RAM. In future, instead of using MCMC posterior sampling, one may use the variational methods in Dirichlet process mixture models to gain more computational efficiency and more scalability as the number of subjects and the number of within-subject measurements increase.

\bibliographystyle{biometrika}
\bibliography{refs}

\begin{thebibliography}{6}
\expandafter\ifx\csname natexlab\endcsname\relax\def\natexlab#1{#1}\fi

\bibitem[{Antoniak(1974)}]{antoniak74}
\textsc{Antoniak, C.~E.} (1974).
\newblock Mixture of {Dirichlet} process with applications to {Bayesian}
  nonparametric problems.
\newblock \textit{Annals of Statistics} \textbf{273(5281)}, 1152--1174.

\bibitem[{Gail(1986)}]{gail1986adjusting}
\textsc{Gail, M.~H.} (1986).
\newblock Adjusting for covariates that have the same distribution in exposed
  and unexposed cohorts.
\newblock \textit{Modern statistical methods in chronic disease epidemiology} ,
  3--18.

\bibitem[{Gail et~al.(1984)Gail, Wieand \& Piantadosi}]{gail1984biased}
\textsc{Gail, M.~H.}, \textsc{Wieand, S.} \& \textsc{Piantadosi, S.} (1984).
\newblock Biased estimates of treatment effect in randomized experiments with
  nonlinear regressions and omitted covariates.
\newblock \textit{Biometrika} \textbf{71}, 431--444.

\bibitem[{Greenland et~al.(1999)Greenland, Robins \&
  Pearl}]{greenland1999confounding}
\textsc{Greenland, S.}, \textsc{Robins, J.~M.} \& \textsc{Pearl, J.} (1999).
\newblock Confounding and collapsibility in causal inference.
\newblock \textit{Statistical Science} , 29--46.

\bibitem[{Lee et~al.(1992)Lee, Wei, Amato \& Leurgans}]{lee1992cox}
\textsc{Lee, E.~W.}, \textsc{Wei, L.}, \textsc{Amato, D.~A.} \&
  \textsc{Leurgans, S.} (1992).
\newblock Cox-type regression analysis for large numbers of small groups of
  correlated failure time observations.
\newblock In \textit{Survival analysis: state of the art}. Springer, pp.
  237--247.

\bibitem[{Sethuraman(1994)}]{sethuraman94}
\textsc{Sethuraman, J.} (1994).
\newblock A constructive definition of dirichlet priors.
\newblock \textit{Statistica Sinica} \textbf{4}, 639--650.

\end{thebibliography}

\end{document}